\documentclass[twocolumn,showpacs,english,superscriptaddress,preprintnumbers,amsmath,amssymb,floatfix]{revtex4-1}

\usepackage[T1]{fontenc}
\usepackage[latin9]{inputenc}
\usepackage{dcolumn}
\usepackage{bm}
\usepackage{graphicx}
\usepackage{color}
\usepackage{esint}
\usepackage{babel}
\usepackage{amsfonts}
\usepackage{slashed}
\usepackage{enumerate}

\begin{document}

\title{Low-energy theory of transport in Majorana wire junctions}

\author{A.~Zazunov}
\affiliation{Institut f\"ur Theoretische Physik, Heinrich-Heine-Universit\"at, D-40225  D\"usseldorf, Germany}

\author{R.~Egger}
\affiliation{Institut f\"ur Theoretische Physik, Heinrich-Heine-Universit\"at, D-40225  D\"usseldorf, Germany}

\author{A.~Levy Yeyati}
\affiliation{Departamento de F{\'i}sica Te{\'o}rica de la Materia Condensada C-V,
Condensed Matter Physics Center (IFIMAC) and Instituto Nicol\'as  Cabrera,
 Universidad Aut{\'o}noma de Madrid, E-28049 Madrid, Spain}

\date{\today}

\begin{abstract}
We formulate and apply a low-energy transport theory for hybrid quantum devices containing junctions of topological superconductor (TS) wires and conventional normal (N) or superconducting (S) leads.  We model TS wires as spinless $p$-wave superconductors and derive their boundary Keldysh Green's function, capturing both the Majorana end state and continuum quasiparticle excitations in a unified manner.  We also specify this Green's function for a finite-length TS wire.  Junctions connecting different parts of the device are described by the standard tunneling Hamiltonian.  Using this Hamiltonian approach, one also has the option to include many-body interactions in a systematic manner. For N-TS junctions, we provide the current-voltage ($I$-$V$) characteristics at arbitrary junction transparency and give exact results for the shot noise power and the excess current.  For TS-TS junctions, analytical results for the thermal noise spectrum and for the $I$-$V$ curve in the high-transparency low-bias regime are presented.   
For S-TS junctions, we compute the entire $I$-$V$ curve and clarify the conditions for having a finite Josephson current.
\end{abstract}


\maketitle

\section{Introduction}\label{sec1}

The physics of topological superconductor (TS) wires, featuring Majorana bound states at their ends, is presently attracting a lot of attention in condensed matter physics, quantum information science, and related fields; for recent reviews, see Refs.~\cite{Alicea2012,Leijnse2012,Beenakker2013,Franz2015,Beenakker2015}. 
Much of this excitement has been fueled by the tremendous experimental progress achieved over the past few years.  Strong evidence for Majorana fermions has 
been reported from transport experiments using topological nanowires proximitized by conventional superconductors \cite{Mourik2012,Das2012,Churchill2013,Albrecht2016} and from scanning tunneling microscopy of magnetic atom chains on  superconducting substrates \cite{Yazdani2014,Franke2015}.  Apart from demonstrating the non-Abelian Majorana braiding statistics, a central goal for future experiments is to thoroughly understand quantum transport in multiterminal hybrid devices containing junctions of TS wires and topologically trivial normal (N) or superconducting (S) materials.  

Problems of this type call for a general and versatile theoretical description capable of treating nonequilibrium transport in such novel devices. One possibility is given by the well-known scattering approach \cite{BTK1982,Nazarov2009}, which has been successfully applied to noninteracting devices containing TS wires \cite{Alicea2012,Leijnse2012,Beenakker2013,Franz2015,Beenakker2015}. We here adapt the Hamiltonian approach \cite{Cuevas1996}, which provides a useful alternative by employing nonequilibrium Green's functions (GFs), to superconducting hybrid systems containing TS wires. This approach starts from uncoupled GFs describing the separate parts of the device, which are then coupled together by tunneling processes. In a noninteracting setting, by solving the Dyson equation, tunnel couplings are taken into account in an exact manner. In addition, by using diagrammatic expansions or related techniques, one can also include many-body interactions. To give just a few examples for successful non-topological applications of the Hamiltonian approach, let us mention S-QD-S  \cite{ALY1997,Zazunov2006} and N-QD-S \cite{Sun1999} junctions containing an interacting quantum dot (QD) sandwiched between S and/or N contacts, extensions to diffusive and/or ferromagnetic systems \cite{Bergeret2005}, Coulomb blockade in voltage-biased superconducting quantum point contacts  \cite{ALY2003,ALY2005}, multiterminal hybrid structures \cite{Melin2002}, and junctions of unconventional superconductors \cite{Datta1998,Sengupta2001,Cuevas2001}. For Majorana wires, similar calculations have been used to describe subgap transport from effective low-energy models that only retain the Majorana sector, see, e.g., Refs.~\cite{Flensberg2010,Leijnse2011,Liu2011}.  

We here derive an explicit and simple expression for the GF describing the boundary of a TS wire, see Eq.~\eqref{negf} below, which captures the Majorana state as well as continuum quasiparticles in a unified manner, and thereby allows for systematic theoretical studies of nonequilibrium transport in topological hybrid devices. We study both subgap and above-gap transport, where detailed and mostly analytical expressions are reported below.  As concrete examples for this approach, we shall here focus on the simplest case given by tunnel junctions. In particular, we discuss the physics of N-TS, TS-TS, and S-TS tunnel junctions involving TS wires with broken time-reversal and spin-rotation symmetries. This ``class $D$'' case is most relevant for present experiments \cite{Mourik2012,Das2012,Churchill2013,Albrecht2016} and corresponds to a spinless $p$-wave superconductor at energies close to the Fermi level \cite{Kitaev2001}.  In more refined descriptions, one could also include high-energy bandstructure effects, see Ref.~\cite{Komnik2015}, order parameter self-consistency, and/or models capturing the phase transition to the non-topological phase. However, analytical results are then generally harder to obtain.  Our theory below allows for arbitrary junction transmission probability $\tau$ (defined for the corresponding N-N junction), bias voltage $V$, and temperature $T$.  
Let us now summarize our main results, explaining also the structure of this paper. 

In Sec.~\ref{sec2}, we present the model and the GF formalism used in this work.  We present the boundary GF both for a semi-infinite and for a finite-length TS wire in Sec.~\ref{sec2a}. In Sec.~\ref{sec2b}, we introduce the tunneling Hamiltonian, followed by  the calculation of transport observables in Sec.~\ref{sec2c}. 

Next, in Sec.~\ref{sec3}, we study transport through a voltage-biased N-TS tunnel junction.  The current-voltage ($I$-$V$) relation for such a junction can always be expressed in terms of a spectral current density $J(\omega)$, which we specify in explicit form in Sec.~\ref{sec3a}. For $\omega=eV$, this spectral density directly determines the $T=0$ differential conductance, for which we arrive at a surprisingly simple result [see Eqs.~\eqref{currentdens} and \eqref{nts-cond} below], valid for arbitrary $\tau$ and $V$.  We thereby reproduce, unify, and simplify previous results \cite{Sengupta2001,Law2009,Flensberg2010,Wimmer2011,Prada2012}. Furthermore, we address the zero-frequency shot noise power in the N-TS junction for voltages below and above the gap, see Sec.~\ref{sec3b}. In the subgap regime, we recover the results of Refs.~\cite{Bolech2007,Nilsson2008,Golub2011} where applicable, while the above-gap results have not been reported elsewhere.  Moreover, we provide closed expressions for the excess current in Sec.~\ref{sec3c}.
 
 In Sec.~\ref{sec4}, we shall discuss TS-TS junctions.  The well-known Josephson effect  for this case \cite{Kitaev2001,FuKane2009,Jiang2011} is briefly discussed within our GF scheme in Sec.~\ref{sec4a}.  In Sec.~\ref{sec4b}, we present analytical expressions for the equilibrium finite-frequency noise spectrum, thereby extending the results of Refs.~\cite{Virtanen2013,Belzig2015} to arbitrary parameters.  Our results also determine the transition rates between Andreev bound states and continuum quasiparticle states. In Sec.~\ref{sec4c}, we study the nonequilibrium multiple Andreev reflection (MAR) features in the time-averaged $I$-$V$ characteristics, cf.~also~Refs.~\cite{Badiane2011,Houzet2013,Aguado2013}, where we provide the excess current and report a closed analytical solution in the large-transparency low-bias regime.    

In Sec.~\ref{sec5}, we study S-TS junctions between a conventional (with gap $\Delta_s$) and a topological  (with gap $\Delta$) superconductor, again for arbitrary junction transparency and arbitrary voltage $V$. In Sec.~\ref{sec5a}, we clarify a recent dispute about the equilibrium Josephson current through such a junction, where Ref.~\cite{Zazunov2012} found a vanishing supercurrent while Ref.~\cite{Ioselevich2015} reported a finite result. We show that tunneling processes have to involve spin flips in 
order to allow for a finite supercurrent in this system.  In Sec.~\ref{sec5b}, we discuss the differential conductance in the absence of spin-flip tunneling processes. We thereby reproduce the recent prediction \cite{Peng2015} of a universal differential conductance peak of height $G_M=(4-\pi) [2e^2/h]$ at $eV=\Delta_s$.  Going beyond Ref.~\cite{Peng2015}, we derive the entire $I$-$V$ curve covering also the above-gap region and parameters away from the tunnel limit.   

We finally offer some conclusions in Sec.~\ref{sec6}.  Details of our calculations can be found in three appendices, and we often employ units with $e= \hbar=k_B=v_F=1$, where $v_F$ is the Fermi velocity. 

\section{Hamiltonian approach}\label{sec2}

\subsection{Green's function formalism}\label{sec2a}

A quantity of central interest for the approach used below is the Keldysh Green's function (GF) $\check{G}$, which is defined for the entire system composed of several tunnel-coupled (super-)conductors. This GF affords a matrix representation on the tensor product of four different spaces:
(i) Keldysh space, referring to the forward/backward parts ($\alpha=+/-$) of the Keldysh time contour needed to properly describe nonequilibrium  transport processes,
(ii) Nambu space encoding the particle/hole structure of the theory, 
(iii) the space labeling different conductors, e.g., the left/right parts ($j=1,2$) of a single tunnel junction, and (iv) time (or frequency) space. 
The structure of $\check{G}$ in Keldysh space, with matrix elements $G^{\alpha \alpha'}$, can be fully expressed in terms of the retarded ($G^R$), advanced $(G^A$), and Keldysh $(G^K)$ GF components \cite{Nazarov2009},
\begin{equation}\label{gfdef}
\check{G}= \left(\begin{array}{cc} G^{++} & G^{+-}\\
G^{-+} & G^{--}  \end{array}\right) =  \check{L} \left( \begin{array}{cc} 0 & G^A  \\ G^R & G^K \end{array} \right) \check{L}^{-1},
\end{equation}
with the Keldysh matrix $\check{L} =\frac{1}{\sqrt{2}} \left(\begin{array}{cc} 1& 1\\
-1 & 1 \end{array}\right)$.

We shall describe the system as built from decoupled pieces that are connected by a tunneling Hamiltonian, cf.~Ref.~\cite{Cuevas1996}.  In such an approach, one first determines the ``uncoupled'' GF $\check{g}$ in the absence of tunnel couplings, which is diagonal in lead space, $\check{g}_{jj'}=\delta_{jj'}\check{g}_{j}$. We shall specify $\check{g}_j$ below for a TS wire ($j=TS$), for a normal conductor ($j=N$), and for a topologically trivial $s$-wave superconductor ($j=S$). In all three cases, it is convenient to use the frequency representation, $\check{g}_j=\check{g}_j(\omega)$.  The Keldysh component $g^K_j(\omega)$, see Eq.~\eqref{gfdef}, is expressed by the retarded/advanced components in a standard manner via the ``local equilibrium'' relation \cite{Nazarov2009},
\begin{equation}\label{kcdef}
g_{j}^K(\omega)= f(\omega) \left( g_{j}^R(\omega)-
g_{j}^A(\omega)\right),
\end{equation}
where the distribution function
\begin{equation}\label{fdef2}
f(\omega) = 1-2n_F(\omega)=\tanh(\omega/2T)
\end{equation} 
is connected to the Fermi function $n_F(\omega)$.   In a gauge (termed ``gauge I'' in what follows) commonly used in the description of normal-conducting systems, tunnel couplings are represented by time-independent matrix elements and one has to take into account the respective chemical potential $\mu_j$ in Eq.~\eqref{kcdef}, see below for details. As is customary for superconducting systems, in  Eq.~\eqref{kcdef} we have instead assumed a different ``gauge II,'' where chemical potential differences appear through time-dependent phases in the tunnel couplings, cf.~Eq.~\eqref{Htun} below.  In any case, once the $\check{g}_j$ are known,  in a second step the full GF $\check{G}$ is obtained by nonperturbatively taking into account tunneling processes via the Dyson equation, see Eq.~\eqref{dyson} below.  From the knowledge of the full GF, all transport quantities of interest can subsequently be determined.  

We begin with the case of a semi-infinite TS wire located at $x>0$, corresponding to lead index $j=TS$. We shall determine the Keldysh GF $\check{g}_{TS}(\omega)$ for electrons/holes near the boundary at $x=0$. The TS nanowire is described as spinless single-channel $p$-wave superconductor, corresponding to the low-energy limit of a Kitaev chain \cite{Alicea2012,Leijnse2012,Kitaev2001}, cf.~Appendix \ref{appa}.  The Hamiltonian reads 
\begin{equation}\label{TSHam}
H_{TS}=\int_0^\infty dx\ \Psi_{TS}^\dagger(x)  \left(-iv_F\partial_x \sigma_z+\Delta\sigma_y\right)\Psi_{TS}(x),
\end{equation}
where the proximity-induced pairing gap $\Delta$ can be chosen real positive.   The Nambu spinor $\Psi_{TS}(x)=(c^{}_r,c^\dagger_l)^T$ in Eq.~\eqref{TSHam} contains right- and left-moving fermion operators $c_{r,l}(x)$, and the Pauli matrices $\sigma_{x,y,z}$ and $\sigma_0={\rm diag}(1,1)$ act in Nambu (particle-hole) space.  
It is well known that the Hamiltonian \eqref{TSHam} corresponds to the 
low-energy form of a generic class-$D$ single-channel TS in one spatial dimension 
\cite{Alicea2012,Leijnse2012,Beenakker2013}.

We emphasize that corrections beyond the ``universal'' class-$D$ 
low-energy model in Eq.~\eqref{TSHam} can be significant for realistic TS 
wires, where the detailed band structure is arguably more complex  
 \cite{Loss2013}.  However, as long as the system remains in 
symmetry class $D$, we expect that predictions based on Eq.~\eqref{TSHam} 
provide at least qualitatively useful answers.  Relying on the topological 
character of the TS phase, one can expect that corrections beyond Eq.~\eqref{TSHam}
allow for a perturbative treatment.  In any case, below we will not discuss such 
corrections, since a decisive advantage of the universal GF, see Eq.~\eqref{negf} 
below, comes from its simplicity and the possibility of obtaining analytical 
results.  Our main goal is not in explaining all possible details of experimental data but rather in providing a unified and coherent theoretical framework, which here will be applied to the simple and widely studied TS wire model \eqref{TSHam}.

The boundary GF $\check{g}_{TS}(\omega)$ can be computed by taking the wide-band limit for a semi-infinite Kitaev chain, or directly by starting from the low-energy Hamiltonian \eqref{TSHam}. We provide a derivation along the first route in App.~\ref{appa}, but one can check that the same result also follows from the second approach.  The GF is defined as the Fourier transform of 
\begin{equation}
\check{g}_{TS}(t-t') = -i \langle {\cal T}_C \Psi(t) \Psi^\dagger(t')\rangle,
\end{equation}
where the boundary Nambu spinor is  $\Psi=(c,c^\dagger)^T$ with $c=[c_l+c_r](x \to 0)$, and ${\cal T}_C$ denotes the Keldysh time-ordering prescription \cite{Nazarov2009}. We note in passing that the relation $\Psi=\sigma_x\Psi^\ast$ (with ``$\ast$'' denoting complex conjugation) imposes a reality constraint on this spinor.
 Using Eqs.~\eqref{gfdef} and \eqref{kcdef},  $\check{g}_{TS}(\omega)$ is fully determined by specifying the Nambu representation of the retarded/advanced GF components, cf.~App.~\ref{appa},
\begin{equation}\label{negf}
g_{TS}^{R/A}(\omega) = \frac{\sqrt{\Delta^2-(\omega\pm i0^+)^2}\ \sigma_0 + \Delta \sigma_x }{\omega\pm i0^+}, 
\end{equation} 
where $R/A$ corresponds to $+/-$, and the branch cut is taken along the negative axis, 
\begin{equation}\label{branchcut}
\sqrt{\Delta^2-(\omega\pm i0^+)^2}= \left\{ \begin{array}{cc} \sqrt{\Delta^2-\omega^2}, &|\omega|\le \Delta, \\
\mp i \ {\rm sgn}(\omega) \sqrt{\omega^2-\Delta^2}, & |\omega|>\Delta.\end{array}\right.
\end{equation}
Below, for retarded (advanced) quantities, the frequency will tacitly be understood as $\omega+ i0^+$ ($\omega-i0^+$). In fact, we shall omit the $R/A$ superscripts whenever the context permits.

From Eq.~\eqref{negf}, the energy-dependent boundary density of states (DOS), $\nu_{TS}(\omega)$, is determined by the Nambu trace of 
\begin{eqnarray}\label{tdos}
 -\frac{1}{\pi} {\rm Im} g_{TS}^R(\omega) &=& \Delta[\sigma_0+\sigma_x]\delta(\omega) \\ \nonumber &+& 
\frac{\sqrt{\omega^2-\Delta^2}}{\pi|\omega|} \sigma_0 \Theta(|\omega|-\Delta)  ,
\end{eqnarray}
with the Heaviside step function $\Theta$, see also Ref.~\cite{Badiane2011}.   Equation \eqref{tdos} features the celebrated Majorana zero-energy peak due to the $\omega=0$ pole of the retarded GF in Eq.~\eqref{negf}.  In addition, for $|\omega|>\Delta$, a continuum quasiparticle contribution is present that vanishes as a square root for $|\omega|\to\Delta$, unlike the conventional BCS singularity, cf.~Eq.~\eqref{bcstdos} below.  The Nambu structure of these two contributions in Eq.~\eqref{tdos} is different and highlights the fact that the Majorana state represents an equal-probability electron-hole superposition state.

We have assumed up to now that the wire is located at $x>0$, where $\check{g}_{TS}=\check{g}_{TS, x>0}$ is evaluated near $x= 0$.  For a wire on the opposite side ($x<0$),  the corresponding boundary GF near $x=0$, $\check{g}_{TS, x<0}$, follows from Eq.~\eqref{negf} by spatial inversion. In effect, due to the $p$-wave character of the superconducting pairing, we need to reverse the sign of $\Delta$ in Eq.~\eqref{negf}, leading to 
\begin{equation}\label{spatialinv}
\check{g}_{TS, x<0}= \sigma_y \check{g}_{TS, x>0}\sigma_y .
\end{equation}
 
Before we specify the corresponding expressions for topologically trivial systems (with $j=N,S$), let us briefly address the effect of a finite TS wire length $L$ on the GF.  In that case, by repeating the analysis in App.~\ref{appa} for a finite-length Kitaev chain with $-L/2\le x\le L/2$, we obtain the retarded/advanced GF near $x=\pm L/2$ as  
\begin{equation}\label{negf2}
g_{TS,  \pm}(\omega) = \frac{\omega \tanh(\zeta_\omega L)}{\omega^2-\epsilon^2_\omega} \left( \zeta_\omega \sigma_0 \mp \tanh(\zeta_{\omega} L) \Delta \sigma_x \right),
\end{equation} 
where $\zeta_\omega=\sqrt{\Delta^2-\omega^2}$ and $\epsilon_\omega=
\Delta/\cosh(\zeta_\omega L)$. Let us show how Eq.~\eqref{negf2} reduces to Eq.~\eqref{negf} in the limit $L\to \infty$. For the subgap part, this is seen in a straightforward manner, but for the continuum spectrum ($|\omega|>\Delta$), one needs to take into account a finite quasiparticle relaxation time $\tau_{qp}$, such that the infinitesimal $0^+$ shift into the complex $\omega$-plane is effectively replaced by $1/\tau_{qp}$. Only then Re$(\zeta_\omega$) is finite and one has $\lim_{L\to \infty}\tanh(\zeta_\omega L)=1$ for frequencies in the continuum part of the spectrum. We note in passing that  Eq.~\eqref{negf2} is also consistent with the spatial inversion rule in Eq.~\eqref{spatialinv}.   On low energy scales, $|\omega|\ll \Delta$, and assuming a long wire with $L>\xi_0$, where $\xi_0=\hbar v_F/\Delta$ is the superconducting coherence length, we conclude that the main finite-$L$ effect in Eq.~\eqref{negf2} is to introduce the hybridization energy scale $\epsilon_\omega\simeq 2\Delta e^{-L/\xi_0}$. This scale describes the exponentially small coupling between the two Majorana end states of a finite-length TS wire.  For $|\omega|>\Delta$, on the other hand, $\zeta_\omega$ becomes imaginary and $\epsilon_\omega$ slowly oscillates with $L$.  In addition, we note that for finite $L$, the off-diagonal (anomalous) part of the GF in Eq.~\eqref{negf2} is suppressed by the last $\tanh(\zeta_\omega L)$ factor.

Consistent with the low-energy TS description, we shall employ the wide-band approximation also in describing  topologically trivial systems. In this standard approximation, the normal density of states is assumed constant near the Fermi level \cite{Nazarov2009}. For a normal metal ($j=N$), the N-TS tunnel coupling effectively involves only one spin component in the normal conductor \cite{Flensberg2010}, and therefore $\check{g}_N$ follows  from Eq.~\eqref{negf}  by letting $\Delta\to 0$,  
\begin{equation}\label{gfn}
g^{R/A}_{N}(\omega) =  \mp i \sigma_0.
\end{equation} 
The corresponding DOS, $\nu_N(\omega)$, is constant. 

For a conventional $s$-wave superconductor ($j=S$) with real positive gap $\Delta_s$, the retarded/advanced GF is given by  \cite{Cuevas1996,Cuevas2001,Avishai2001}, 
\begin{equation}\label{gfs}
g_{S}(\omega) = -\frac{\omega\sigma_0 + \Delta_s \sigma_x}{\sqrt{\Delta_s^2-\omega^2}}, 
\end{equation}
resulting in the familiar DOS of a BCS superconductor. The latter is proportional to 
\begin{equation}\label{bcstdos} 
\nu_S(\omega) = \frac{|\omega|}{\sqrt{\omega^2-\Delta_s^2}} \Theta(|\omega|-\Delta_s).
\end{equation} 
In that case, Nambu spinors of the boundary fields are defined as $\Psi_S=(c_\uparrow,c_\downarrow^\dagger)^T$, where the spinful fermion operator $c_{\uparrow/\downarrow}=c_{l,\uparrow/\downarrow}+c_{r,\uparrow/\downarrow}$ is given by the sum of the left- and right-moving components.  

\subsection{Tunneling Hamiltonian}\label{sec2b}

We now include the tunneling Hamiltonian $H_T$ connecting different conductors. For the moment, we shall employ gauge II, cf.~Sec.~\ref{sec2a}, where chemical potential differences enter through time-dependent phase factors in $H_T$. 

Let us start with a single tunnel junction,  leaving aside the $j=S$ case discussed later on.  Using operators $c_{j=1,2}$ for electrons close to the left/right side of the junction, the standard tunneling Hamiltonian reads 
\begin{equation}\label{Htun}
H_T(t)=  \lambda e^{i\phi(t)/2} c_1^\dagger c_2^{} + {\rm h.c.}
\end{equation}
Without loss of generality, the hopping amplitude $\lambda$ is assumed real-valued. The normal transmission probability $\tau$ of the junction (with $0\le \tau\le 1$) is then given by \cite{Nazarov2009,Cuevas1996}
\begin{equation}\label{taudef}
\tau = 4\lambda^2/(1+\lambda^2)^2 .
\end{equation}
An applied d.c.~bias voltage, $eV=\mu_1-\mu_2$, appears here through the phase difference 
\begin{equation}\label{josephsonrel}
\phi(t) = [\phi_1-\phi_2](t)= \phi_0 + 2eVt/\hbar.
\end{equation}
In the equilibrium case ($V=0$), only the static phase $\phi_0$ is present.  We note that with our unit conventions and normalization of surface GFs, the tunnel coupling $\lambda$ in Eq.~\eqref{Htun} implicitly includes density-of-states factors due to the leads, containing in particular the different Fermi velocities of the (super-)conductors on both sides of the junction.  (This statement also applies to the S-TS case.)

It is convenient to express Eq.~\eqref{Htun} in Nambu representation, where we also generalize the formalism to an arbitrary number $M$ of conductors, $j=1,\ldots, M$.  
For that purpose, we first define the time-dependent tunneling matrix $W(t)$. In lead space, all diagonal elements of $W$ vanish, $W_{jj}=0$, while the off-diagonal elements are given by the Nambu matrices (here, $j< j'$)
\begin{equation} \label{TDef}
W^{}_{jj'}(t) = \lambda_{jj'} \sigma_z e^{i\sigma_z[\phi_j(t)-\phi_{j'}(t)]/2}, \quad 
W_{j'j}^{}(t)=W_{jj'}^\dagger(t).
\end{equation}  
The tunneling Hamiltonian then follows in the form 
\begin{equation} \label{Htun1}
H_T(t) =\frac12 \sum_{jj'}^M \Psi_j^\dagger W_{jj'}(t) \Psi_{j'}^{}, \quad \Psi_{j}=\left(\begin{array}{c} c_{j}\\ c_{j}^\dagger\end{array}\right),
\end{equation}
and the Heisenberg operator describing the current flowing through lead $j$ is given by
\begin{equation}\label{currentoper}
\hat I_j(t) = \frac{2e}{\hbar} \frac{\delta H_T(t)}{\delta \phi_j(t)} =
i\sum_{j'\ne j} \Psi_j^\dagger(t) \sigma_z W_{jj'}(t) \Psi_{j'}^{}(t) .
\end{equation}

We now discuss how to describe S-TS junctions, putting for simplicity $M=2$. The Nambu spinor on the left ($j=S$) BCS superconducting side is $\Psi_1=(c_{1,\uparrow}^{},c_{1,\downarrow}^\dagger)^T$, and in the absence of spin-flip tunneling, $H_T$ is given in general form as
\begin{equation}
H_T(t) =  \lambda e^{i\phi(t)/2} [ \cos(\theta)c_{1,\uparrow}^\dagger+e^{-i\chi}\sin(\theta) c_{1,\downarrow}^{\dagger}] c_2^{} + {\rm h.c.},
\end{equation}
with two additional real-valued parameters $\chi$ and $\theta$ on top of the gauge-invariant phase difference $\phi(t)$.  
The junction transparency is again expressed in terms of $\lambda$ by Eq.~\eqref{taudef}. Performing the gauge transformation
$c_{1,\uparrow/\downarrow}\to e^{\pm i\chi/2} c_{1,\uparrow/\downarrow}$, the phase $\chi$ can be absorbed by renormalizing the static phase difference $\phi_0\to \phi_0+\chi$ in Eq.~\eqref{josephsonrel}.
We can therefore put $\chi=0$ in what follows.  In addition, by exploiting the SU$(2)$ spin symmetry of the $s$-wave BCS superconductor, we may also put $\theta=0$, again without loss of generality \cite{footnote2}.  Written in Nambu notation, $H_T$ is then as in Eq.~\eqref{Htun1}, where instead of Eq.~\eqref{TDef}, $W(t)$ has the non-zero Nambu matrix element
\begin{equation}\label{WSTS}
W_{12}(t) = \lambda e^{i\phi(t)/2} \Pi_\uparrow,\quad \Pi_\uparrow =\frac{\sigma_0+\sigma_z}{2},
\end{equation}
with $W_{21}(t)=W_{12}^\dagger(t)$. In this basis, due to the presence of the projection operator $\Pi_\uparrow$ in Eq.~\eqref{WSTS}, only spin-$\uparrow$ electrons in the BCS superconductor are tunnel-coupled to the effectively spinless fermions on the $j=TS$ side. Such a spin-filtered tunnel coupling is generic for junctions without spin-flip tunneling.  For example, if the junction contains magnetic impurities, this property will be lost and the theory has to be modified.

\subsection{Transport observables}\label{sec2c}

In the absence of many-body interactions, by using the equations of motion for Heisenberg operators, we obtain the ``full'' Keldysh GF as solution of the Dyson equation
\begin{equation}\label{dyson}
\check{G} = \left(\check{g}^{-1} - \check{W}\right)^{-1},
\end{equation}
with the Keldysh matrix $\check{W}={\rm diag}(W,-W)$.
From this solution, all nonequilibrium transport quantities of interest can be deduced as described next. In addition, many-body interactions can be included by well-established perturbative/diagrammatic techniques \cite{Nazarov2009}.

Let us first discuss the mean current flowing through the $j$th lead, $I_j(t)$, which in general will be time-dependent.  Taking the expectation value of the current operator  \eqref{currentoper}, $I_j$ is expressed in terms of the Keldysh GF component ($G^K$) at coinciding times,  
\begin{equation}\label{currentgeneral}
I_j(t) = \frac12 \sum_{j' \ne j} {\rm tr}_N \left( \sigma_z W_{jj'}(t) G^K_{j'j}(t,t)\right),
\end{equation}
where the trace ``tr$_N$'' is over Nambu space, and  current conservation dictates the condition $\sum_j I_j(t)=0$.  In order to evaluate $G^K$, we now employ Eqs.~\eqref{kcdef} and \eqref{dyson}.  For arbitrary gauge,  we find  
\begin{equation}\label{hhh}
G^K = G^R  F -  F G^A + G^R ( F W - W F ) G^A,
\end{equation}
where matrix products correspond to convolutions  and $F_{jj'}=\delta_{jj'}F_j$ contains the distribution functions in the absence of tunneling. Explicitly, in gauge II,  $F_j(\omega)=f(\omega)\sigma_0$ with $f(\omega)$ in Eq.~\eqref{fdef2}. In gauge I, on the other hand, for time-independent chemical potential and a normal-conducting system, one finds 
\begin{equation}\label{fdef3}
F_j(\omega) =  \left(\begin{array}{cc} f(\omega-\mu_j) & 0\\ 0 & f(\omega+\mu_j)\end{array}\right),
\end{equation}
which can be rationalized by noting that the upper (lower) entry describes electrons (holes).  

Next, we turn to the current-current correlation function (``noise''),
\begin{equation}\label{noisedef}
S_{jj'}(t,t') = \left\langle \delta \hat I_j(t) \delta \hat I_{j'}(t')\right \rangle,\quad \delta \hat I_j(t)=\hat I_j(t)-I_j(t), 
\end{equation} 
which can similarly be expressed in terms of the full GF. Using the Keldysh GF components $G^{+-}$ and $G^{-+}$, cf.~Eq.~\eqref{gfdef}, these noise correlations follow as
\begin{widetext}
\begin{equation}\label{noisegeneral}
 S_{jj'}(t,t') = \sum_{j_1\ne j}^M \sum_{j_2\ne j'}^M \ {\rm tr}_N\left ( \sigma_z W_{jj_1}(t) \left [ G_{j_1j_2}^{-+}(t,t') \sigma_z W_{j_2j'}(t') G_{j'j}^{+-}(t',t) - G_{j_1j'}^{-+}(t,t') \sigma_z W_{j'j_2}(t') G^{+-}_{j_2j}(t',t)\right]\right ).
\end{equation}
\end{widetext}

To give a first example for the above expressions, the time-averaged current-voltage characteristics of a tunnel junction ($M=2$) between an arbitrary pair of the above systems follows at low transparency, $\tau\ll 1$, from a lowest-order perturbative solution of the Dyson equation \eqref{dyson}.   Equation \eqref{currentgeneral} 
thereby yields the current $I=I_1=-I_2$ as
\begin{equation}\label{tunnel-current}
I(V) =   \frac{e\tau}{2h} \int d\omega \ \nu_1(\omega) \nu_2(\omega-eV) \left[ f(\omega)-f(\omega-eV) \right],
\end{equation}
with the energy-dependent DOS $\nu_{1,2}(\omega)$ on the respective side, and $f(\omega)$ in Eq.~\eqref{fdef2}.

\section{N-TS junction}\label{sec3}

In this section, we shall study a tunnel junction between a normal conductor  ($j=1$) and a TS wire ($j=2$).   Going beyond Eq.~\eqref{tunnel-current}, we consider the case of arbitrary junction transparency $0\le \tau\le 1$.   For an N-TS junction at constant bias voltage, $eV=\mu_1-\mu_2$, it is convenient to adopt gauge I with time-independent tunneling matrix $W_{12}=\lambda \sigma_z$.

\subsection{Differential conductance}\label{sec3a}

The current-voltage characteristics of the N-TS junction follows from Eq.~\eqref{currentgeneral} as
\begin{equation} \label{ntscurr}
I(V) = \frac{\lambda}{2} \int \frac{d\omega}{2\pi} {\rm tr}_N  G^K_{21}(\omega) ,
\end{equation}
where Eq.~\eqref{hhh} determines the needed Keldysh GF component,
\begin{eqnarray} \label{gknts}
&& G^K_{21}(\omega) = G^R_{21} F_1-F_2 G^A_{21} + \\ &+&  \lambda \nonumber
\left [ G^R_{21}\sigma_z \left(F_1-F_2\right)
G^A_{21}  - G^R_{22}\sigma_z \left(F_1-F_2\right) G^A_{11} \right],
\end{eqnarray}
with the distribution functions 
\begin{equation}
 F_1(\omega)=f(\omega-V \sigma_z)\sigma_0, \quad F_2(\omega) = f(\omega)\sigma_0.
\end{equation}
The retarded/advanced GF components appearing in Eq.~\eqref{gknts} are obtained by solving the Dyson equation (\ref{dyson}),
\begin{eqnarray}\nonumber
G_{11}(\omega) &=& \left(\left [g_1(\omega)\right]^{-1} - \lambda^2 \sigma_z g_{2}(\omega) \sigma_z \right)^{-1},\\  \nonumber
G_{22}(\omega) &=& \left(\left [g_2(\omega)\right]^{-1} - \lambda^2 \sigma_z g_{1}(\omega) \sigma_z \right)^{-1},
\\  \label{greensfunctions}
G_{21}(\omega) &=& \lambda g_2(\omega) \sigma_z G_{11}(\omega).
\end{eqnarray}
The uncoupled GF $g_1$ for the normal part is given by Eq.~\eqref{gfn}, and the TS counterpart $g_2$ by Eq.~\eqref{negf}.   

Inserting Eq.~\eqref{gknts} into Eq.~\eqref{ntscurr}, we obtain the current-voltage characteristics,
\begin{equation}\label{currents}
 I = \frac{e}{h} \int d\omega\left[ n_F(\omega-eV)-n_F(\omega+eV) \right]  J(\omega),
\end{equation}
and the differential conductance
\begin{equation}\label{condnts}
G = \frac{dI}{dV} =  \frac{2e^2}{h} \int_{-\infty}^\infty d\omega  \frac{J(\omega)}{4T\cosh^2[(\omega-eV)/2T]} .
\end{equation}
The spectral current density is symmetric, $J(\omega)=J(-\omega)$, and follows in remarkably simple form,
\begin{equation}\label{currentdens}
J(\omega) = \left\{\begin{array}{cc} 1/(1+\omega^2/\Gamma^2),& |\omega|<\Delta,\\ & \\
  \tau \frac{\tau+(2-\tau)\sqrt{1-(\Delta/\omega)^2}}{\left[2-\tau+\tau\sqrt{1-(\Delta/\omega)^2}\right]^2} ,& |\omega|\ge \Delta,
\end{array} \right.
\end{equation}
with the rate \cite{Tanaka2000}
\begin{equation}\label{ratedef}
\Gamma= \frac{ \tau\Delta}{2\sqrt{1-\tau}}.
\end{equation}
Note that $J(\omega)$ remains continuous at $\omega\to \Delta$, where $J(\Delta)=\tau^2/(2-\tau)^2.$   

In the subgap regime $|\omega|<\Delta$, Eq.~\eqref{currentdens} yields a Lorentzian peak of width $\Gamma$ centered around $\omega=0$, which describes the Majorana bound state leaking into the normal  conductor with hybridization $\Gamma$.  For $\tau\ll 1$, the above-gap part of the spectral density is given by 
\begin{equation}\label{abovegapsmalltau}
J_{\tau\ll 1}(|\omega|>\Delta) \simeq \frac{ \tau}{2}\sqrt{1-\Delta^2/\omega^2},
\end{equation}
which provides only a subleading contribution to the conductance for low junction transparency.
On the other hand, in the limit of a fully transparent junction with $\tau=1$, the rate $\Gamma$ diverges and Eq.~\eqref{currentdens} reduces to 
\begin{equation} \label{currhight}
J_{\tau=1}(\omega) = \left \{ \begin{array}{cc} 1, & |\omega|<\Delta,% \\  &
 \\
\left(1+\sqrt{1-\Delta^2/\omega^2}\right)^{-1}, & |\omega|\ge  \Delta.
\end{array}\right.
\end{equation}

\begin{figure}[t]
 \begin{center}
\includegraphics[width=9.1cm]{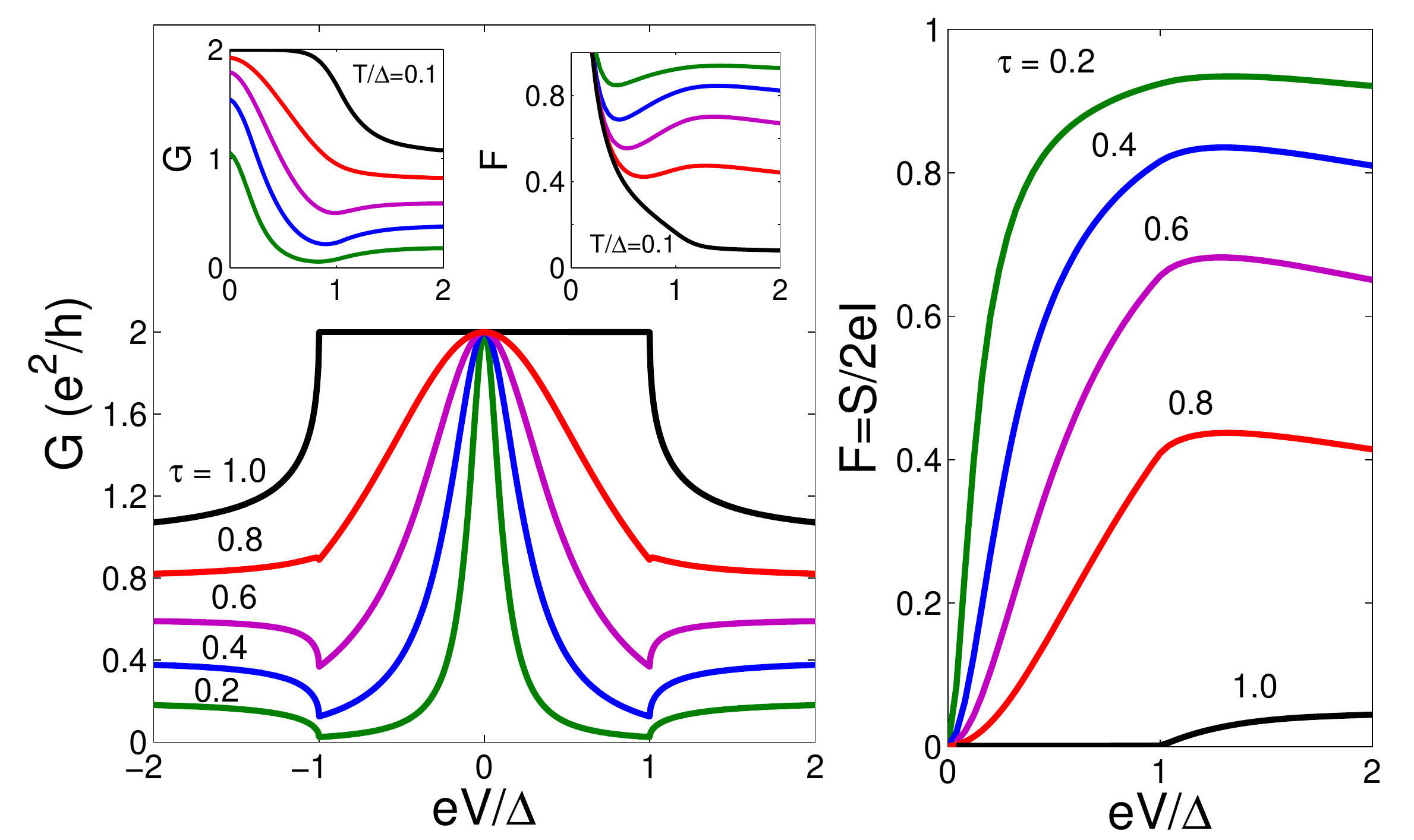}
\end{center}  
\caption{ (Color online) Transport observables of the N-TS junction. Left panel: The main part shows the differential conductance $G=dI/dV$ vs $eV/\Delta$, see Eq.~\eqref{nts-cond}, for several transmission probabilities $\tau$ and $T=0$.  The left inset shows corresponding results for $T=0.1\Delta$, the right inset shows the Fano factor for the same temperature. Right panel: Fano factor $F=S/(2eI)$ vs $eV/\Delta$, see Eq.~\eqref{nts-noise}, for the same values of $\tau$ and $T=0$.} \label{f1}
\end{figure}

Let us now discuss the differential conductance, see Eq.~(\ref{condnts}), in the most interesting zero-temperature limit, where
\begin{equation}\label{nts-cond}
G(V,T=0) =  \frac{2e^2}{h} J(eV)
\end{equation}
is directly expressed in terms of the spectral current density $J(\omega)$ in Eq.~\eqref{currentdens}.  Equation \eqref{nts-cond} is equivalent to a more complicated expression reported in Ref.~\cite{Sengupta2001}. It recovers the celebrated Majorana zero-bias peak with quantized peak height $2e^2/h$ and width $\Gamma$ due to resonant Andreev reflection  \cite{Bolech2007,Law2009,Flensberg2010}.  Near perfect transmission, $\tau\to 1$,  although the Majorana state is not well-defined anymore in view of the strong N-TS hybridization, conductance quantization still remains robust \cite{Sengupta2001,Wimmer2011}. In fact, $G=2e^2/h$ persists throughout the entire subgap regime $|eV|<\Delta$, see Eq.~\eqref{currhight}. For $|eV|\gg \Delta$, Eq.~\eqref{nts-cond} approaches the Ohmic conductance $\tau e^2/h$ expected for a normal-conducting spinless tunnel junction in the classical regime.  These results are illustrated in the left panel of Fig.~\ref{f1}.

The finite-temperature behavior of the conductance  can be 
analyzed in a similar manner.  The subgap Lorentzian peak in  $J(\omega)$, cf.~Eq.~\eqref{currentdens}, causes the finite-$T$ lineshape of a conventional resonant tunneling conductance peak \cite{Alicea2012,Nazarov2009}, featuring a temperature-induced decrease (increase) of the peak conductance (width) as illustrated in the left inset 
in Fig.~\ref{f1}.

\subsection{Shot noise}\label{sec3b}

In addition to the conductance, another transport property of interest is the shot noise power, i.e., the Fourier transformed current-current correlation function,
\begin{equation}\label{noiseLL}
S_{11}(\omega)=-S_{12}(\omega) =  \int dt e^{i\omega (t-t')} S_{11}(t,t'),
\end{equation}
where $S_{11}(t,t')$ has been defined in Eq.~\eqref{noisedef}. Representing the correlation function by GFs, see Eq.~\eqref{noisegeneral}, one obtains an integral representation for $S_{11}(\omega)$. We here study the shot noise power in the zero-frequency limit, $S(V) = 2S_{11}(\omega\to 0)$,
which is compared to its Poissonian value  $2eI(V)$ \cite{Nazarov2009}. 

Our results for the Fano factor $F=S/(2eI)$ at temperature $T=0$ are shown in the right panel of Fig.~\ref{f1}.  We observe that for $\tau\ll 1$, the Poissonian limit $F=1$ as predicted in Ref.~\cite{Bolech2007} is approached.   With increasing transparency and/or lower bias voltage, however, $F$ is reduced and ultimately vanishes in the entire subgap regime at perfect transparency ($\tau=1$).  In fact, for $|eV|<\Delta$, we reproduce the analytical $T=0$ result of Ref.~\cite{Golub2011},
\begin{equation}\label{nts-noise}
S = \frac{4e^2\Gamma}{h} \left ( \tan^{-1}(eV/\Gamma) - \frac{eV/\Gamma}{1+(eV/\Gamma)^2} \right) ,
\end{equation}
with the rate $\Gamma$ in Eq.~\eqref{ratedef}. The corresponding Fano factor $F=S/(2eI)$, with $I=(2e\Gamma/h)\tan^{-1}(eV/\Gamma)$, perfectly fits the subgap part of the results shown in the right panel of Fig.~\ref{f1}. 

Beyond reproducing Eq.~\eqref{nts-noise} for the subgap regime, the GF approach also yields the shot noise power for voltages above the gap. For $eV\gg \Delta$, the Fano factor approaches the value $F=1-\tau$, which is expected for the corresponding spinless N-N junction \cite{Nazarov2009}.  We note that even in the limit of ideal transparency ($\tau=1$), the above-gap $T=0$ shot noise is finite due to the simultaneous presence of both Andreev and quasiparticle processes \cite{Khlus1987}.

Finally, results for the finite-temperature Fano factor are displayed in the right inset of the left panel in Fig.~\ref{f1}.  The strong thermal component in the noise power, 
$S(V=~0)=4TG(V=0)=8Te^2/h$, now completely dominates the $V\to 0$ behavior and leads to an upturn of all curves as the voltage is reduced.   

\subsection{Excess current}\label{sec3c}

We conclude our study of the N-TS junction with a discussion of the excess current,
which can be directly measured in experiments and is defined as
\begin{equation}\label{excdef}
 I_{\rm exc}=\lim_{V \to \infty} \left[ I(V) - I_{\rm N-N}(V) \right],
\end{equation} 
where $I_{\rm N-N}=\tau e^2 V/h$ is the normal-state ($\Delta=0)$ current for the same junction.
For $e V > \Delta$, Eq.~\eqref{currents} gives the $T=0$ current for the N-TS junction in the form
\begin{eqnarray}\label{nts:Ints}
I_{\rm N-TS}(V) &=&  \frac {e\tau \Delta}{ h}  \frac{\tan^{-1}(2 \sqrt{1-\tau}/{\tau})}{ \sqrt{1-\tau}}  \\ \nonumber &+&\frac {2 e w\Delta}{ h}
\int_1^{e V / \Delta} dx \, \frac{ x \left( w x + \sqrt{x^2 - 1} \right)}{ \left( x + w \sqrt{x^2 - 1} \right)^2 } ,
\end{eqnarray}
with $w =\tau/(2-\tau)$. The first (voltage-independent) term is a subgap contribution to the total current, while the second term comes from quasiparticles with energies above the superconducting gap.

The integral in Eq.~\eqref{nts:Ints} can be rationalized by Euler's substitution $t = x + \sqrt{x^2 - 1}$.
Performing the integration over the new variable $t$ and using an asymptotic expansion in $\Delta / eV$, the excess current follows in closed form as
\begin{widetext}
\begin{equation}\label{excnts}
I_{\rm exc, N-TS} =  \frac{e\tau \Delta}{ h} \left[ \frac{\tan^{-1}\left(\frac{2 \sqrt{1-\tau}}{\tau}\right)}{ \sqrt{1-\tau}}   +(1-\tau)^{-3/2} \left\{\tau\sqrt{1-\tau} - \left( 1 + (1-\tau)^2 \right) \left[\frac{\pi}{2}- \tan^{-1}\left (\frac{1} {\sqrt{1-\tau}}\right)  \right]\right\} \right] .
\end{equation}
\end{widetext}
The excess current \eqref{excnts} is always positive. In particular, for $\tau = 1$, one obtains $I_{\rm exc, N-TS} = (4/3) (e \Delta / h)$, which is half the value of the excess current, $I_{\rm exc, N-S} = (8/3) (e \Delta / h)$,
in a conventional (topologically trivial) ballistic N-S contact with full transparency \cite{Cuevas1996,Zaitsev1980}.
For $\tau<1$, the relative suppression factor is slightly less than $1/2$.

\section{TS-TS junction}\label{sec4}

Next we turn to the case of a TS-TS junction. For clarity, we shall assume identical absolute values of the pairing gap on both sides, $\Delta_1=\Delta_2=\Delta$. In Secs.~\ref{sec4a} and \ref{sec4b}, we discuss the equilibrium case ($V=0$), where the Josephson junction is biased by a static phase difference $\phi_0$, and the tunnel matrix $W$ in Eq.~\eqref{TDef} has non-zero elements $W_{12}=W_{21}^\dagger=\lambda \sigma_z e^{i\phi_0\sigma_z/2}$.  We subsequently turn to the voltage-biased case in Sec.~\ref{sec4c}.

\subsection{Fractional Josephson effect}\label{sec4a}

  As detailed in App.~\ref{appb}, using a similar calculation as in the N-TS case, Eq.~\eqref{currentgeneral} yields the  equilibrium Josephson current-phase relation in the form  
\begin{eqnarray}\label{fracjos1}
I(\phi_0) &=& -\frac{e\tau}{4\hbar}\Delta^2 \sin(\phi_0) \int \frac{d\omega}{2\pi i} f(\omega) \\
&\times& \left( \frac{1}{(\omega+i0^+)^2-E_A^2}
- \frac{1}{(\omega-i0^+)^2-E_A^2} \right),\nonumber
\end{eqnarray}
where $f(\omega)$ is given by Eq.~\eqref{fdef2} and we define the Andreev bound state energy 
\begin{equation}\label{andreevstate}
E_A(\phi_0) = \sqrt{\tau} \Delta \cos(\phi_0/2).
\end{equation}
The  integral in Eq.~\eqref{fracjos1} can be done by residues, with poles at $\omega=\pm E_A$ infinitesimally shifted into the complex plane. The  $\pm$ sign corresponds to the conserved fermion parity eigenvalue of the Josephson junction, cf.~Ref.~\cite{Alicea2012} for a detailed discussion, and generates a pair of decoupled (i.e., crossing) $4\pi$-periodic Andreev bound states with dispersion $\pm E_A(\phi_0$).   From Eq.~\eqref{fracjos1},  we obtain
\begin{equation}\label{fracjos}
I(\phi_0) = \frac{e\sqrt{\tau}\Delta}{2\hbar} \sin(\phi_0/2) \tanh(E_A/T) ,  
\end{equation}
without contributions from continuum quasiparticles.
Since the GF formalism implicitly assumes a thermodynamic average, Eq.~\eqref{fracjos} represents an average over both parity states. The resulting current-phase relation is therefore $2\pi$-periodic. By restricting the integral in Eq.~\eqref{fracjos} to a specific parity eigenvalue, one may arrive at the well-known ``fractional'' Josephson effect with a $4\pi$-periodic current-phase relation  \cite{Kitaev2001,FuKane2009,Jiang2011} instead of Eq.~\eqref{fracjos}.  Parity conservation is more directly visible in 
our study of noise properties in Sec.~\ref{sec4b}, where it is responsible for the 
absence of transitions within the Andreev bound state sector. 

\subsection{Thermal finite-frequency noise}\label{sec4b}

Next we discuss the (unsymmetrized) current noise at finite frequency, $S_+(\omega)=S_{11}(\omega)$, see Eq.~\eqref{noiseLL}, where we 
consider the equilibrium case allowing for analytical progress. (Nonequilibrium aspects of quantum noise in TS-TS junctions have been studied in Refs.~\cite{Badiane2011,Houzet2013}.) Putting $V=0$,  Eq.~\eqref{noisegeneral} yields the thermal noise correlations in the form (cf.~App.~\ref{appb})
\begin{eqnarray}\label{thermalnoise1}
S_{+}(\omega) &=& \frac{e^2}{h} \int d\omega_1d\omega_2 \delta(\omega_1-\omega_2+\omega) \\ &\times& 
\nonumber n_F(\omega_1) [1-n_F(\omega_2)] Q(\omega_1,\omega_2).
\end{eqnarray} 
Here, $Q=Q_{A-c}+Q_{c-c}$ is symmetric in the frequency arguments, $Q(\omega_1,\omega_2)=Q(\omega_2,\omega_1)$, and can be decomposed into a part $Q_{A-c}$ due to transitions between the Andreev bound state sector  (with $|\omega|=| E_A|)$ and continuum quasiparticle states (with $|\omega|>\Delta$), plus a continuum contribution $Q_{c-c}$.  
However, there is no contribution from the Andreev sector alone, i.e., $Q_{A-A}=0$. This result should be contrasted to the case of non-topological S-S junctions, where transitions at frequency $\omega=2E_A$ are always present for $\tau<1$ \cite{ALY1996} and imply $Q_{A-A}\ne 0$. The  absence of direct transitions between the two Andreev bound states in a TS-TS junction can be understood as manifestation of fermion parity conservation, cf.~Refs.~\cite{FuKane2009,Badiane2011,Virtanen2013,Belzig2015}.  Technically, in our approach, $Q_{A-A}=0$ can be traced back to the orthogonality of different current eigenstates. While current eigenstates always coincide with Andreev bound states for TS-TS junctions, this holds true only at perfect transmission ($\tau=1$) for the S-S case \cite{AZ2005}.

\begin{figure}[t]
\begin{center} 
\includegraphics[width=9.5cm]{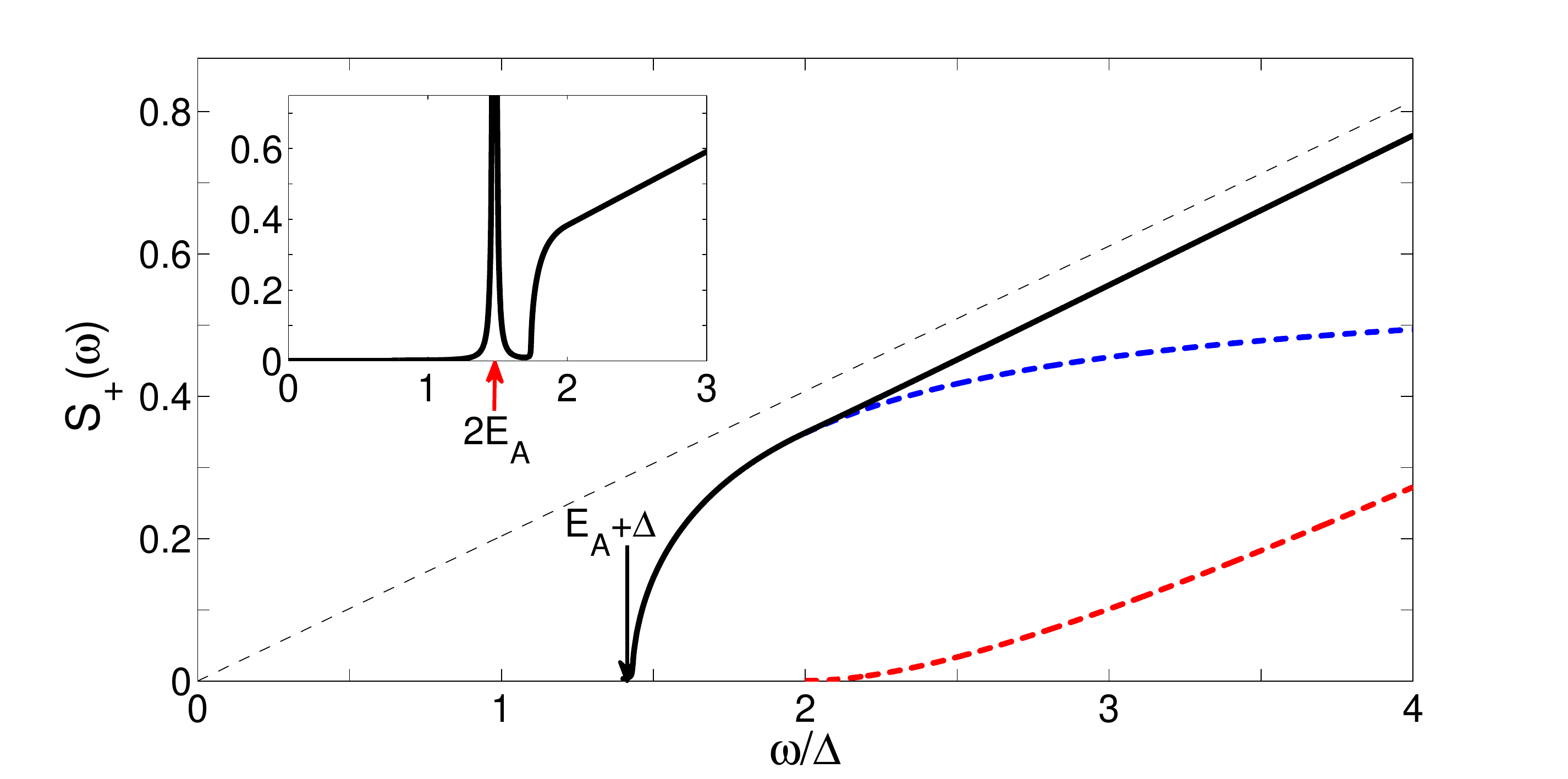}
 \end{center}  
\caption{ (Color online) Finite-frequency noise spectrum $S_+(\omega)$ vs $\omega/\Delta$ for $T=V=0$ with $\tau=0.64$ and $\phi_0=2$, where $S_+$ is given in units of $e^2\Delta/\hbar$.  The main panel is for the TS-TS case, where the blue dashed curve gives $S_{A-c}$ in Eq.~\eqref{Acnoise}, the red dashed curve gives $S_{c-c}$ from Eq.~\eqref{ccnoise}, and the solid black curve shows $S_+=S_{A-c}+S_{c-c}$. In the shown frequency range, $S_{c-c}$ is well approximated by Eq.~\eqref{omegalow}. The thin-dashed black curve gives the $\Delta =0$ result, i.e., the leading term in Eq.~\eqref{omegainf}. The inset shows $S_+(\omega)$ for a topologically trivial S-S junction with the same parameters, see Ref.~\cite{ALY1996}, where the $\omega=2E_A$ peak with $E_A^{(S-S)}(\phi_0) =  \Delta\sqrt{1-\tau\sin^2(\phi_0/2)}$ has been broadened by replacing the infinitesimal shift $0^+\to 0.001\Delta$ in the GFs.   }\label{f2}
\end{figure}

As is shown in App.~\ref{appb}, Andreev-continuum transitions yield
\begin{eqnarray}\label{QAc}
&& Q_{A-c}(\omega_1,\omega_2)  = \pi \tau  
\delta(|\omega_1|-|E_A|) \Theta(|\omega_2|-\Delta) \\
\nonumber & &\qquad \times \, \frac{\sqrt{(\Delta^2-\omega_1^2)(\omega_2^2-\Delta^2)}}
{|\omega_2|-{\rm sgn}(\omega_1\omega_2) |\omega_1|} + (\omega_1\leftrightarrow 
\omega_2),
\end{eqnarray}
while the continuum part $Q_{c-c}$ involves both inter- and intra-band transitions, 
\begin{eqnarray}\label{Qcc}
&& Q_{c-c}(\omega_1,\omega_2) =2\tau \Theta(|\omega_1|-\Delta)\Theta(|\omega_2|-\Delta) \\ \nonumber &&\times \, \frac{\sqrt{(\omega_1^2-\Delta^2) (\omega_2^2-\Delta^2)}}{(\omega_1^2-E_A^2) (\omega_2^2-E_A^2) } \left( |\omega_1\omega_2|+{\rm sgn}(\omega_1\omega_2) E_A^2 \right).
\end{eqnarray}
The finite-frequency noise (\ref{thermalnoise1}) thus receives two contributions, $S_+(\omega)=S_{A-c}(\omega)+S_{c-c}(\omega)$.
Let us now discuss these two contributions to $S_+(\omega)$ at $T=0$, cf.~Fig.~\ref{f2}.

In the zero-temperature limit, the Andreev-continuum contribution follows from Eqs.~\eqref{thermalnoise1} and
\eqref{QAc} in the form
\begin{eqnarray}\label{Acnoise}
S_{A-c}(\omega) &=& \frac{e^2\tau}{\hbar} \sqrt{\Delta^2-E_A^2}
\\ &\times& \nonumber \Theta\left(\omega-|E_A|-\Delta\right) 
\frac{\sqrt{(\omega-|E_A|)^2-\Delta^2}}{\omega},
\end{eqnarray}
and is finite only for $\omega>\Delta + |E_A|$. The continuum contribution requires $\omega>2\Delta$, where we find
\begin{eqnarray}\label{ccnoise}
S_{c-c}(\omega) &=& \frac{2e^2\tau}{\hbar}  \Theta(\omega-2\Delta)
\int_{\Delta-\omega}^{-\Delta} d\omega_1\\ &\times&
 \sqrt{(\omega_1^2-\Delta^2)[(\omega_1+\omega)^2-\Delta^2 ]} \nonumber \\
 \nonumber &\times& 
 \frac{-\omega_1(\omega_1+\omega)-E_A^2}{(\omega_1^2-E_A^2)[(\omega_1+\omega)^2-E_A^2]}.
\end{eqnarray}
For frequencies near the threshold, $\omega-2\Delta\ll \Delta$, this gives
\begin{equation}\label{omegalow}
S_{c-c}( \omega)  \simeq \frac{e^2   \tau }{4 \hbar} 
 \frac{\Delta(\omega-2\Delta)^2}{\Delta^2 - E_A^2}\Theta(\omega-2\Delta),
\end{equation}
while for $\omega\gg \Delta$, Eq.~\eqref{ccnoise} yields
\begin{equation}\label{omegainf}
S_{c-c}(\omega) \simeq \frac{e^2\tau}{\pi\hbar} \left[\omega - \sqrt{\Delta^2-E_A^2} \tan^{-1}\left( \sqrt{\frac{\omega^2 - \Delta^2 }{ \Delta^2 - E_A^2} }\right)\right].
\end{equation}
The above results are illustrated in Fig.~\ref{f2}. We note that the fluctuation-dissipation theorem relates the frequency-dependent admittance of the junction to $S_+(\omega)$.  In particular, transition rates between Andreev  and continuum states directly follow from Eq.~\eqref{Acnoise}, cf.~Eq.~(14) in Ref.~\cite{Kos2013}. 
In the inset of Fig.~\ref{f2}, we compare the above results to finite-frequency noise in a topologically trivial S-S junction with otherwise identical parameters \cite{ALY1996}.  Clearly, for the TS-TS junction, there is no $\omega=2E_A$ peak, and the frequency dependence of $S_+(\omega)$ for $\omega>\Delta+|E_A|$ is rather different. 

\subsection{Current-voltage characteristics}\label{sec4c}

We now turn to the case of a voltage-biased TS-TS junction, where we shall discuss the time-averaged current-voltage characteristics in the $T=0$ limit.  As is well known, subgap transport is then governed by MAR processes.  We here briefly show that our approach recovers previous results \cite{Badiane2011,Houzet2013,Aguado2013}, and then point out that the low-bias regime admits an analytical solution.

For a numerical evaluation of the current-voltage characteristics, it is convenient to adopt gauge II in the Hamiltonian description. One can then follow the strategy discussed in Ref.~\cite{Cuevas1996}, where the corresponding problem has been solved for voltage-biased S-S contacts. For the TS-TS case, we can similarly expand the mean current as
$I(t) = \sum_n \tilde I_n e^{in\omega_0t}$ with $\omega_0 = 2eV/\hbar$, where we arrive at expressions relating the current coefficients $\tilde I_m$ to double Fourier GF components 
($\check{G}_{nm}$) formally identical to the expressions in Ref.~\cite{Cuevas1996}. The recursive algorithm devised in Ref.~\cite{Cuevas1996} then directly applies after replacing the uncoupled GFs by $\check{g}_{TS}$, cf.~Eq.~\eqref{negf}, and yields the numerically exact solution for the time-dependent current flowing through the junction. 

\begin{figure}[t]
\begin{center} 
\includegraphics[width=9.5cm]{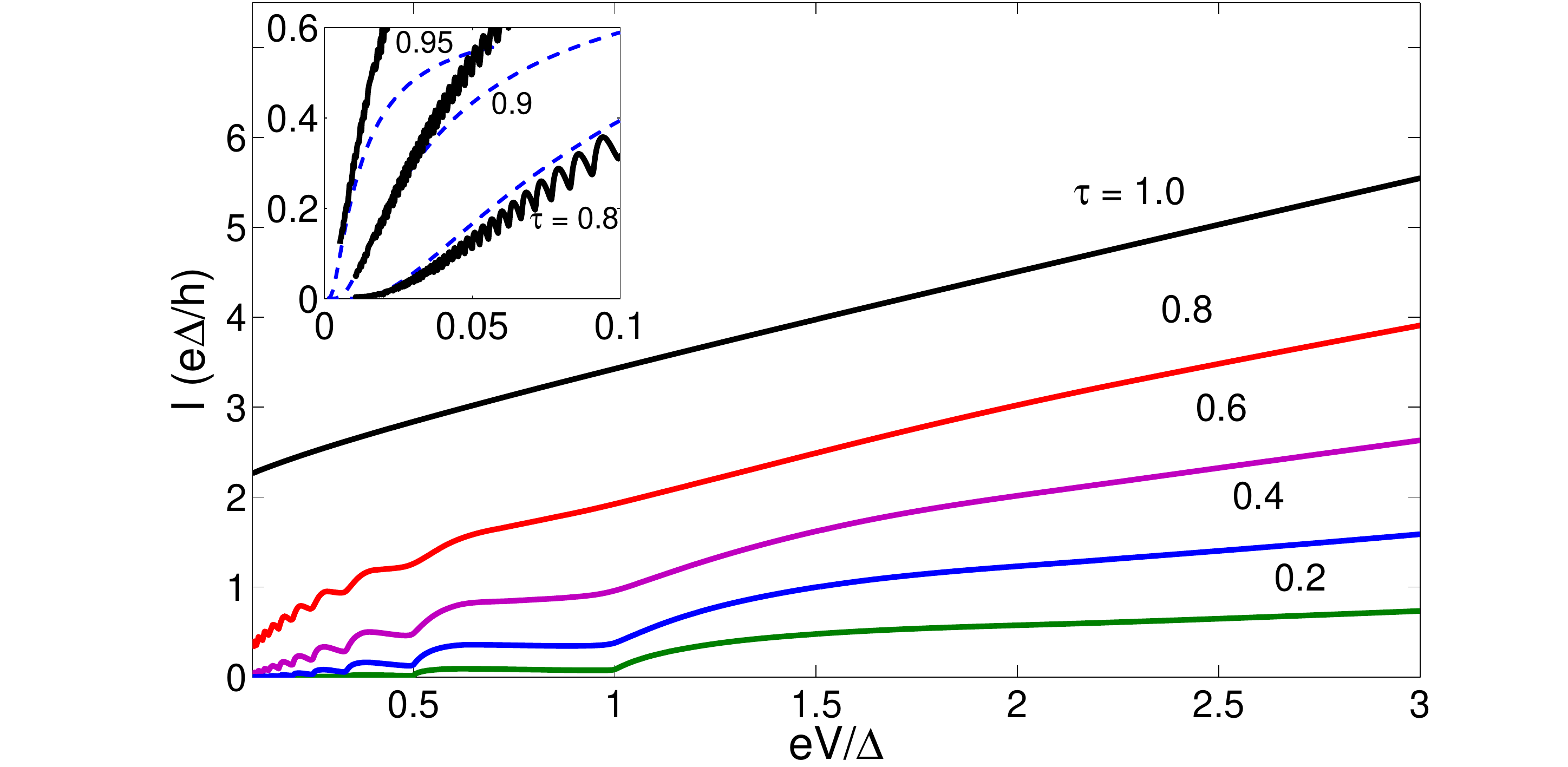}
 \end{center}  
\caption{ (Color online) Time-averaged current-voltage characteristics $I$ vs $V$ of a TS-TS junction at zero temperature for several transparencies $\tau$. Inset: Same for the low-bias regime, $eV\ll \Delta$, with high transparency. The blue dashed curve shows the analytical prediction in Eq.~\eqref{mar2003}, while the solid curves follow from numerically exact calculations.  }\label{f3}
\end{figure}

In Fig.~\ref{f3}, we show the resulting d.c.~component $I=\tilde I_0$ as a function of the bias voltage for several  junction transparencies $\tau$.  We find that the current exhibits subgap steps at $eV = 2\Delta/(2n)$ with integer $n$, which are more and more rounded as $\tau$ increases. These steps correspond to the onset of even-order Andreev reflection processes. We note that in a conventional S-QD-S junction containing a resonant dot state, such subgap steps happen at $eV=2\Delta/(2n+1)$ \cite{ALY1997}, i.e., only odd-order Andreev reflection processes contribute. For the TS-TS case at hand, as explained in Ref.~\cite{Aguado2013}, the opposite situation is encountered and only even orders are important.  
Eventually, at perfect transmission ($\tau=1$), a practically linear dependence on voltage is reached after an abrupt conductance jump to $2e^2/h$ at zero bias. Our results in Fig.~\ref{f3} agree with those of Ref.~\cite{Badiane2011} obtained by employing scattering theory for the time-dependent Bogoliubov-de Gennes equation.

As has been discussed, e.g., in Ref.~\cite{Cuevas1996}, the low-bias behavior of a superconducting junction in the MAR regime can be understood in terms of the dynamics of Andreev bound states. We here exploit the close relation between low-bias transport in a TS-TS junction and for an S-QD-S contact, where the tunnel junction contains an interacting quantum dot at resonance. The latter problem has been analyzed in Ref.~\cite{ALY2003}, where the Andreev bound state spectrum is well approximated by 
$E_A \simeq \tilde{\Delta} \cos(\phi_0/2)$, with a renormalized amplitude $\tilde{\Delta} < \Delta$.  The Andreev bound state dispersion is formally identical to the TS-TS junction case in Eq.~\eqref{andreevstate} with the identification $\tilde\Delta=\sqrt{\tau}\Delta$. However, while the $4\pi$-periodicity of the Andreev states is robust and protected by parity conservation for the TS-TS junction, it is only accidental in the S-QD-S case, since taking into account asymmetries in left/right tunnel couplings and/or shifting the dot level slightly away from resonance, a gap opens and $2\pi$-periodicity will be restored.  In particular, for the S-QD-S case, spin degeneracy results in four possible states, with the ``even'' sector corresponding to the $\pm E_A$ states and the ``odd'' sector to a pair of spin-degenerate zero-energy states.  In spite of these subtleties, this correspondence yields an analytical solution for the low-bias ($eV\ll \Delta$) part of the $I$-$V$ curve in the TS-TS junction. With the rate $\Gamma^*= \tau\Delta / 4\sqrt{1-\tau}$, which is precisely one-half of the N-TS rate $\Gamma$ in Eq.~\eqref{ratedef}, we obtain
\begin{equation}\label{mar2003}
I(V) = \frac{2e}{\Gamma^* V\hbar} \int_\Delta^\infty dx x^2\sqrt{\frac{x^2-\Delta^2}{x^2-\tilde\Delta^2}} e^{2x(\tanh\alpha -\alpha)/V},
\end{equation}
where $\cosh\alpha=x/\tilde\Delta$, see Ref.~\cite{ALY2003}.  This analytical result is shown as dashed line in the inset of Fig.~\ref{f3} and well describes our numerical results  in the limit $eV\ll \Delta$ and $\tau\agt 0.8$.  We mention in passing that Eq.~\eqref{mar2003} also agrees well with the analytical approximation in Ref.~\cite{Houzet2013} for sufficiently low bias voltage, even though their expression looks
rather different.

We conclude this section by noting that the excess current \eqref{excdef} for a TS-TS junction with gaps $\Delta_1$ and $\Delta_2$ is given by
\begin{equation} \label{exctsts}
I_{\rm exc, TS-TS}=I_{\rm exc, N-TS}(\Delta_1)+I_{\rm exc,N-TS}(\Delta_2),
\end{equation}
where the $T=0$ excess current of an N-TS  junction with the same transparency $\tau$  has been specified in Eq.~\eqref{excnts}.  Technically, this result follows by noting that only single Andreev reflection processes survive for $V\to \infty$ \cite{Cuevas1996}.
For equal gaps, Eq.~\eqref{exctsts} predicts a doubling of the TS-TS excess current relative to the corresponding N-TS value.

\section{S-TS junction}\label{sec5}

In this section, we study the current $I$ flowing through an S-TS junction between a conventional $s$-wave BCS superconductor on the left side ($j=1$) and a topological TS wire on the right side ($j=2$). The respective gaps are denoted by $\Delta_1=\Delta_s$ and $\Delta_2=\Delta$. The noise properties can also be determined using the present GF formalism, see Eq.~\eqref{noisegeneral}, but we leave this question to future work. 
  
As discussed in Sec.~\ref{sec2b}, we here consider spin-conserving tunneling processes, where (after a suitable basis choice) only spin-$\uparrow$ fermions in the $s$-wave superconductor are tunnel-coupled to the effectively spinless TS wire. The tunneling matrix $W$ follows from  
$W_{12}=\lambda e^{i\phi(t)/2} \Pi_\uparrow$, see Eq.~\eqref{WSTS},
with the projection operator $\Pi_\uparrow=(\sigma_0+  \sigma_z)/2$. The time-dependent mean current flowing through the junction can be computed from the general expression in Eq.~\eqref{currentgeneral}. 
Working in gauge II and using the Dyson equation \eqref{dyson}, we find  
\begin{eqnarray}\label{currsts}
I(t) &=& -  \lambda^2 \ {\rm Re}  \int dt'  e^{-i[\phi(t)-\phi(t')]/2}
\\  \nonumber &\times& {\rm tr}_N
\left[ \tilde g^{R}_1(t-t') G_{22}^K(t',t) +\tilde g^K_1(t-t') G_{22}^A(t',t) \right].  
\end{eqnarray}
Here, projected GFs for the $s$-wave superconductor are defined  by
\begin{equation}\label{projgf}
\tilde g^{R/A/K}_1(t)=\Pi_\uparrow g^{R/A/K}_S(t)  \Pi_\uparrow,
\end{equation} 
with the Fourier transform $\check g_S(t)$ of $\check g_S(\omega)$ in Eq.~\eqref{gfs}. For details on the derivation of Eq.~\eqref{currsts}, see App.~\ref{appc}.

\subsection{Equilibrium S-TS Josephson current} \label{sec5a}

The equilibrium Josephson current through a phase-biased S-TS junction has previously been studied by two of us \cite{Zazunov2012}, where we found that there are no Andreev bound  states and hence the Josephson current vanishes identically, $I(\phi_0)=0$, as long as tunneling remains spin-conserving.  This result finds a simple explanation by noting the different pairing symmetries on both sides of the junction: their orthogonality effectively blocks the supercurrent.   In fact, in the absence of spin flips during tunneling events, the present GF approach confirms this result explicitly from Eq.~\eqref{currsts} after putting $V=0$, as we briefly demonstrate in App.~\ref{appc}.  

However,  recent theoretical work \cite{Ioselevich2015} reported a finite Josephson current through an S-TS junction, where the $s$-wave superconductor has been represented by two (opposite-spin) Kitaev chains in the continuum limit.  Employing a scattering approach under the assumption of full channel mixing at the junction, which implicitly requires strong spin-flip scattering, the Josephson current was then shown to be finite.  Our approach can easily handle spin-flip scattering during tunneling \cite{Cuevas2001}, and we have reproduced the results of Ref.~\cite{Ioselevich2015} by such a generalization.  However, we here refrain from a detailed discussion of this issue, and instead continue with the $I$-$V$ characteristics of an S-TS junction under the assumption of spin-conserving tunneling. This case is encountered, for instance, when electrons/holes are tunneling from a superconducting scanning tunneling microscope tip  through vacuum to the edge of a TS wire.

\subsection{Voltage-biased S-TS junction} \label{sec5b}

Next we turn to a discussion of the time-averaged current through a voltage-biased S-TS junction. For a constant voltage bias, we have $\phi(t)=2eVt/\hbar$,  and the d.c.~current $I(V)$ through the S-TS junction follows from Eq.~\eqref{currsts} after some algebra given in App.~\ref{appc}.  For the same reason that Andreev bound states do not appear in the equilibrium case, MAR phenomena are absent in this setup. We therefore do not need a double Fourier representation of the GF.  Despite of this simplification, the result given below is a bit lengthy, but at the same time it is exact for arbitrary parameter values.  We note in passing that the conventional superconductor 
is here assumed to be tunnel-coupled to the edge of the TS wire.  When the junction is instead located some distance $d$ away from the edge, one has to evaluate the GF
 $g_{TS}(x,x';\omega)$ at position $x=x'=d$,  cf.~Ref.~\cite{Peng2015}. The latter GF can be computed using similar steps as given in App.~\ref{appa}. 

Using  $\omega_\pm= \omega\pm eV$ and $f(\omega)=1-2n_F(\omega)$ in Eq.~\eqref{fdef2},  Eq.~\eqref{currsts} yields
\begin{eqnarray}\label{stsnv}
I(V) &=&\lambda^2 {\rm Re}\int \frac{d\omega}{2\pi}\Bigl[ \gamma_1^R(\omega_-) G_{22;ee}^K(\omega)
\\ \nonumber &+& 2i f(\omega_-)\nu_1(\omega_-) G^A_{22;ee}(\omega) \Bigr],
\end{eqnarray}
where $G_{22;ee}$ refers to the $(1,1)$ entry of the corresponding Nambu matrix with 
\begin{equation}
G^{R/A}_{22;ee}(\omega) = \frac{-\gamma_{2}(\omega) z_h(\omega)}{z_e(\omega) z_h(\omega)- 1/(1-\omega^2/\Delta^2) }
\end{equation}
and
\begin{eqnarray}\nonumber
G_{22;ee}^K(\omega) &=& \frac{-2i\left |\gamma_{2}^R(\omega)\right|^2 }{\left|
z_e^A(\omega) z_h^A (\omega) - 1/[1-(\omega-i0^+)^2/\Delta^2]\right|^2}\\ 
 &\times& \left( f_e(\omega)\left|z_h^R(\omega)\right|^2+\frac{f_h(\omega) }{|1-\omega^2/\Delta^2|} \right).
\end{eqnarray}
We use the DOS $\nu_1(\omega)$ of the BCS superconductor in Eq.~\eqref{bcstdos}, and the continuum part of the DOS of the TS wire, cf.~Eq.~\eqref{tdos}, $\nu_{2}(\omega)=\Theta(|\omega|-\Delta)\sqrt{1-\Delta^2/\omega^2}.$
In addition, we introduce effective distribution functions $f_{e/h}(\omega)$ for electrons ($e$) and holes ($h$), 
\begin{equation}\label{feh}
f_{e/h}(\omega) = f(\omega) \nu_{2}(\omega) + \lambda^2 f(\omega_\mp) \nu_1(\omega_\mp).
\end{equation}
Since $f(0)=0$, the Majorana peak in the DOS of the TS wire does not contribute to $f_{e/h}(\omega)$, and only the continuum part of $\nu_{2}(\omega)$ matters here.  Finally, in the above expressions,  we employ the retarded/advanced quantities
\begin{eqnarray}
\gamma_{j}(\omega) &=& \frac{\omega}{\sqrt{\Delta_{j}^2-\omega^2}},\\ \nonumber
z_{e/h}(\omega) &=& 1-\lambda^2\gamma_1(\omega_\mp)\gamma_2(\omega). 
\end{eqnarray}

In the limit $\Delta_s\to 0$, one finds after a short calculation that Eq.~(\ref{stsnv}) reduces to the current through an N-TS junction, see Eq.~\eqref{currents}.
Furthermore, for $\tau\ll 1$, Eq.~\eqref{stsnv} reduces to the $I$-$V$ relation in Eq.~\eqref{tunnel-current} applicable in the deep tunneling regime.
An interesting recent study \cite{Peng2015} for precisely the same S-TS junction has reported a universal peak height of the differential low-temperature conductance. This conductance peak is asymmetric and sets in at $eV=\Delta_s$, where $G$ jumps to the value  
\begin{equation}\label{PengG}
G_M = (4-\pi) \frac{2e^2}{h}.
\end{equation}
Such a feature may be useful for the detection of Majorana bound states.  Equation~\eqref{PengG} has been derived by projecting away the TS continuum quasiparticles, i.e., by formally sending $\Delta\to\infty$ \cite{Peng2015}.  Indeed, in that case, Eq.~\eqref{stsnv} simplifies to 
\begin{equation}\label{ifelix}
I= \frac{4e}{h}\int d\omega \left(n_F(\omega_-)-n_F(\omega_+)\right) \frac{\nu_1(\omega_-)\nu_1(\omega_+)}{\left[\nu_1(\omega_-)+\nu_1(\omega_+)\right]^2} .
\end{equation}
At low temperatures, $T\ll \Delta_s$, and for voltages $eV=\Delta_s+\eta$ with $|\eta|\ll \Delta_s$, where the BCS singularity in the $s$-wave superconductor lines up 
with the Majorana zero-energy level at $\eta=0$, Eq.~\eqref{ifelix} yields
\begin{eqnarray}\nonumber
 I&=& \frac{8e}{h} \Theta(\eta) \int_{0}^\eta d\omega\frac{\frac{1}{\sqrt{\eta^2-\omega^2}}}{\left( \frac{1}{\sqrt{\eta+\omega}}+\frac{1}{\sqrt{\eta-\omega}}\right)^2}\\
 &=& (4-\pi)\frac{2e}{h}(eV-\Delta_s) \Theta(eV-\Delta_s) ,
\end{eqnarray}
directly leading to Eq.~\eqref{PengG}.  Note that no current flows for $eV<\Delta_s$. 

\begin{figure}[t]
\begin{center} 
\includegraphics[width=9.5cm]{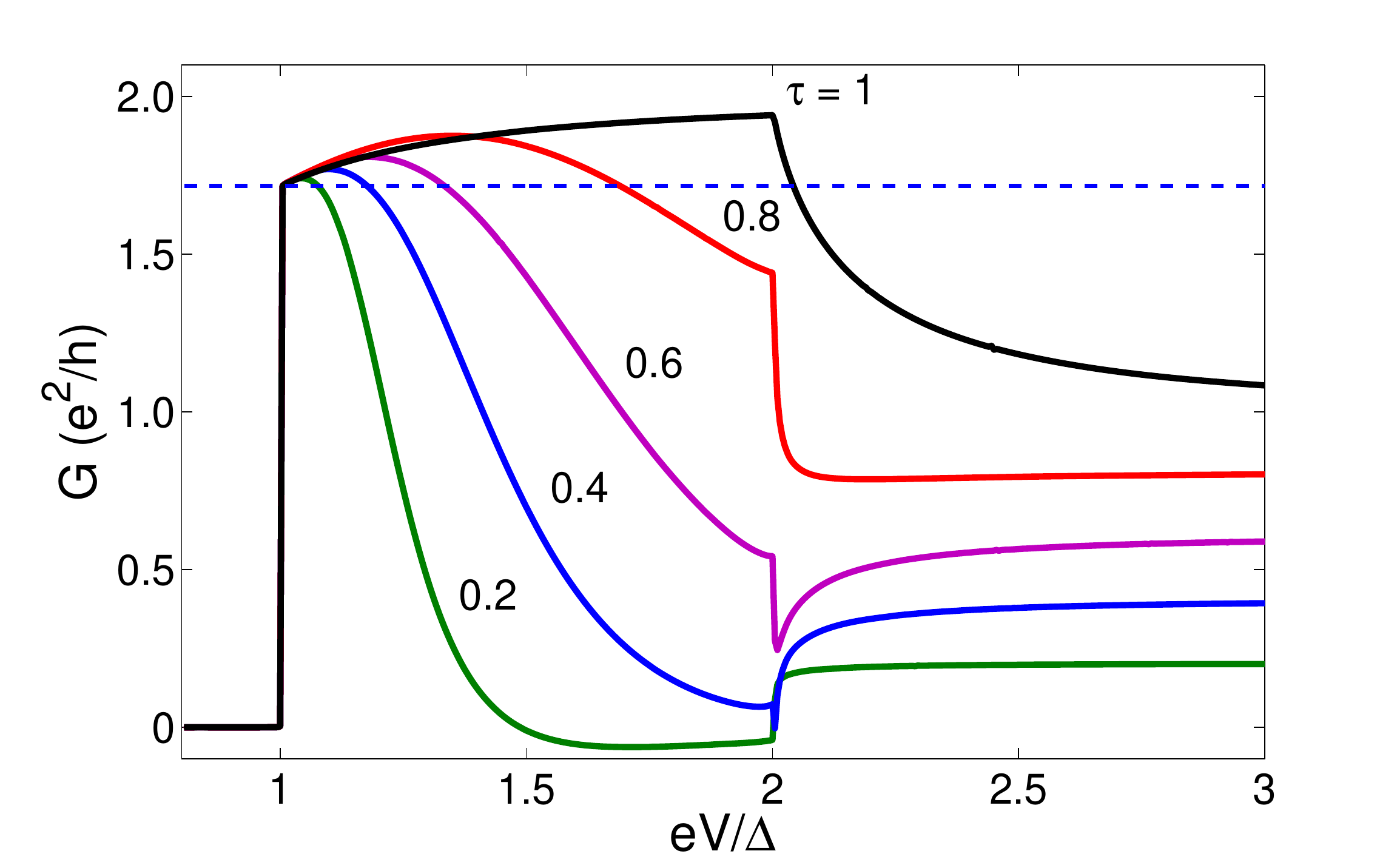}
 \end{center}  
\caption{ (Color online) Differential conductance $G=dI/dV$ vs $V$ at zero temperature for an  S-TS junction with $\Delta_s=\Delta$ and several transparencies $\tau$. For $|eV|<\Delta$, the current vanishes. The dashed blue line gives $G_M$ in Eq.~\eqref{PengG}. }\label{f4}
\end{figure}

We illustrate the differential conductance $G=dI/dV$ obtained from Eq.~\eqref{stsnv} in Fig.~\ref{f4}, where $G_M$ is indicated by the dashed blue line.  The universal peak height (\ref{PengG}) follows for arbitrary $\tau$ as $eV$ approaches $\Delta_s$ from above.  For $\tau\ll 1$, Fig.~\ref{f4} also confirms the subgap conductance lineshape, i.e, the dependence on $V$ for $eV<\Delta+\Delta_s$, derived in Ref.~\cite{Peng2015}.

Generally, we observe from Fig.~\ref{f4} that the conductance first increases with increasing voltage, and then strongly decreases except for very high junction transparency. Note in particular that the conductance can become negative, see Fig.~\ref{f4}. Such a negative differential conductance is not surprising when tunneling through a bound state, cf.~Ref.~\cite{Andersen2011}. For $\tau=1$, the ideal resonant Andreev reflection value $G=2e^2/h$ is (almost) reached as $eV\to\Delta+\Delta_s$. For voltages $eV>\Delta+\Delta_s$ and $\tau=1$, the conductance then drops in a continuous fashion.  However, for $\tau<1$, we find that the conductance exhibits a finite jump to a smaller value as the voltage goes through this threshold value separating the subgap from the above-gap regime.  Furthermore, at very large voltage, the conductance again approaches the Ohmic value $\tau e^2/h$ of the corresponding N-N junction.
We emphasize that finite temperature effects are exponentially small due to the presence of a gap on both sides of the junction.

\section{Conclusions and outlook}\label{sec6}

To conclude, we have formulated a general nonequilibrium Green's function framework to study transport in hybrid devices containing Majorana wires.  Our approach employs the
boundary Green's function of such a Majorana wire, given in Eq.~\eqref{negf}, which is sufficiently simple to allow us to derive several new analytical results and/or provide
expressions that can be treated numerically in a straightforward manner.  As applications, we have discussed three elementary tunnel junctions involving topologically nontrivial wires,  
where we take into account both the Majorana sector and the continuum quasiparticles on equal footing.

There are many interesting applications that could be treated  in the future by this Hamiltonian approach. For instance, the approach should be suitable to study multiterminal junctions or networks containing TS wires \cite{Alicea2011,Weithofer2014,Tarasinski2015,Haim2015}, the coupling of Majorana wires to (interacting) quantum dots \cite{Leijnse2011,Liu2011}, and/or when a finite-length TS wire is contacted by several electrodes \cite{Nilsson2008,DasSarma2012}. Other possible directions are to include a.c.~voltages in order to study, e.g., fractional Shapiro steps in TS-TS junctions \cite{Jiang2011,Badiane2011,Houzet2013,Aguado2012},  or to study the interplay between Coulomb charging effects \cite{Fu2010,Zazunov2011,BeriCooper2012,Huetzen2012}
and the presence of continuum quasiparticles. Moreover, it will be interesting to extend the boundary GF given above for the class-$D$ wire also to other symmetry classes as well as to topological superconductors of dimensionality higher than one. 
A related generalization may employ a GF for the $p$-wave superconductor covering a wider parameter regime, such that one can study the phase transition between the non-topological and the topological phase. We leave those extensions for future work.

\acknowledgments  
We wish to thank R. Aguado, K. Flensberg, T. Jonckheere, T. Martin, A. Mart{\'i}n-Rodero, J. Rech, and P. Recher for helpful discussions.  This work has been supported by the Deutsche Forschungsgemeinschaft (Bonn) within Grant No.~EG 96-10/1, and by the Spanish MINECO through Grant No.~FIS2014-55486-P
and through the ``Mar\'{\i}a de Maeztu'' Programme for Units of Excellence in R\&D (MDM-2014-0377). 

\appendix
\section{Boundary GF of semi-infinite TS wire }\label{appa}

Here we provide a derivation of the retarded/advanced GF $g_{TS}^{R/A}(\omega)$ in Eq.~\eqref{negf}, which describes the low-energy dynamics of electrons and holes at the boundary of a semi-infinite TS wire.  Let us start from the standard Kitaev chain model \cite{Kitaev2001} in the topologically nontrivial phase, for simplicity with chemical potential $\mu=0$. Using the pairing amplitude $\Delta$ (assumed real positive), the hopping matrix element $t_0$, and spinless fermion operator $c_x$ for lattice site no.~$x$ (with lattice spacing $a=1$), the ``bulk''  Hamiltonian is 
\begin{eqnarray}\label{kitaevchain}
H^{(b)}_K &=& \frac12  \sum_x   \left(- t_0 c_x^\dagger c_{x+1}^{} + \Delta c_x^{} c_{x+1}^{}\right) + {\rm h.c.}  \\ \nonumber
&=& \frac12 \sum_k \Psi_k^\dagger h_k \Psi_k,\quad h_k=-t_0 \cos(k) \sigma_z+\Delta\sin(k) \sigma_y, 
\end{eqnarray}   
with Nambu spinors $\Psi_k=(\psi_k,\psi_{-k}^\dagger)^T$ subject to the reality constraint $\Psi^{}_k=\sigma_x\Psi_{-k}^\ast$ and Pauli matrices $\sigma_{x,y,z}$ in Nambu space.  In Eq.~\eqref{kitaevchain}, we assume periodic boundary conditions, $c_{x+N}=c_x$, and write
\begin{equation}
\Psi(x) = \left(\begin{array}{c} c_x \\ c_x^\dagger\end{array} \right) = \frac{1}{\sqrt{N}}
\sum_k e^{ikx} \Psi_k,
\end{equation}
with the number of lattice sites $N\to \infty$.  We note in passing that a linearization of the Hamiltonian (\ref{kitaevchain}) around the two Fermi points, $k_F=\pm \pi/2$ (half-filling), obtains
\begin{equation}
 H_K^{(b)} \simeq \sum_q \Phi_q^\dagger \left( v_F q \sigma_z +\Delta \sigma_y\right)\Phi_q ,\quad \Phi_q= \left( \begin{array}{c} \psi_{\pi/2+q}\\ \psi^\dagger_{-\pi/2-q}
\end{array}\right),
\end{equation}
with Fermi velocity $v_F=t_0$, see Eq.~\eqref{TSHam}.
 
The ``bulk'' retarded/advanced GF of $\Psi(x)$ for the translationally invariant Kitaev chain in Eq.~(\ref{kitaevchain}) is given by the Nambu matrix 
\begin{equation}\label{gf1}
g^{(b)}_{xx'}(\omega) =\frac{1}{N}\sum_k \left(\omega-h_k\right)^{-1} e^{ik(x-x')}.
\end{equation}
Passing to the continuum representation in momentum space, it is convenient to evaluate Eq.~\eqref{gf1} as a sum of residues in the $z=e^{ik}$ plane,
\begin{equation}\label{residues1}
g^{(b)}_{xx'}(\omega) = \oint_{|z|=1} \frac{dz}{2\pi i} (\omega-h_k)^{-1} 
z^{x-x'-1},
\end{equation}
where we only need the result for lattice sites $x,x'\in \{ 0,\pm 1\}$ below.
From Eq.~\eqref{residues1}, we find  
\begin{eqnarray}\label{g111}
 g^{(b)}_{00}(\omega) &=& \frac{-2\omega\sigma_0}{\sqrt{(\omega^2- \Delta^2)(\omega^2-4t_0^2)}},
\\ \nonumber
g^{(b)}_{\pm 1,0}(\omega) &=& g^{(b)}_{0,\mp 1}(\omega) =  
\frac{2t_0(z_1^2+1)\sigma_z \pm  i \Delta(z_1^2-1)\sigma_y}
{\sqrt{(\omega^2-\Delta^2)(\omega^2-4t_0^2)}},
\end{eqnarray}
where 
\begin{eqnarray*}
z_1^2 &=& \frac{2\omega^2-(4t_0^2+\Delta^2)}{4t_0^2-\Delta^2} - {\rm sgn}\left( 2 \omega^2-(4t_0^2+\Delta^2)\right) \\ &\times& 
\sqrt{\left(\frac{2\omega^2-(4t_0^2+\Delta^2)}{4t_0^2-\Delta^2}\right)^2-1}.
\end{eqnarray*}
In the wide-band limit defined by $t_0\gg {\rm max}(\Delta,|\omega|)$ \cite{Nazarov2009}, Eq.~\eqref{g111} simplifies to
\begin{eqnarray} \label{g112}
 g^{(b)}_{00}(\omega) &=& \frac{-\omega}{t_0 \sqrt{\Delta^2-\omega^2}}\sigma_0,\\ g^{(b)}_{\pm 1,0}(\omega) &=& g^{(b)}_{0,\mp 1}(\omega) \nonumber
= \frac{\sqrt{\Delta^2-\omega^2}\sigma_z\mp i\Delta \sigma_y}{t_0\sqrt{\Delta^2-\omega^2}}.
\end{eqnarray}

In the next step, we add a local potential scatterer of strength $U$ at site $x=0$, resulting in the Hamiltonian $H_K= H^{(b)}_K+Uc_0^\dagger c^{}_0.$
The ``full'' GF, $g_{xx'}(\omega)$, then obeys the Dyson equation
$g= g^{(b)} + g^{(b)} U\sigma_z g$.
Letting the impurity strength $U\to \infty$, one effectively cuts the chain at site $x=0$, and therefore the boundary GF of the semi-infinite TS wire follows as 
$g_{TS}(\omega)=g_{11}(\omega)$. Solving the above Dyson equation for  $U\to \infty$, we obtain, see also Ref.~\cite{arrachea},
\begin{equation} \label{boundarygf}
g_{TS}(\omega) =  g^{(b)}_{00}(\omega) - g^{(b)}_{10}(\omega)
\left[ g^{(b)}_{00}(\omega)\right]^{-1} g^{(b)}_{01}(\omega).
\end{equation}
By inserting the wide-band expression \eqref{g112} for $g_{xx'}^{(b)}(\omega)$ into Eq.~\eqref{boundarygf}, we finally arrive at $g_{TS}^{R/A}(\omega)$ as quoted in  Eq.~\eqref{negf}, see Sec.~\ref{sec2}.

\section{On TS-TS junctions}\label{appb}

In this Appendix, we provide derivations for several of our results on TS-TS junctions in Sec.~\ref{sec4}. 

First, let us sketch how to obtain the Josephson current-phase relation $I(\phi_0)$ in  Sec.~\ref{sec4a}. By using Eq.~\eqref{currentgeneral}
and adapting the expressions in Secs.~\ref{sec2} and \ref{sec3} from the N-TS to the TS-TS case, we arrive at the integral representation
\begin{equation}\label{integraliphi0}
I(\phi_0) = \frac12\int \frac{d\omega}{2\pi} f(\omega) {\rm tr}_N\left( \sigma_z \left[ X^R(\omega)-X^A(\omega) \right] \right).
\end{equation}
Here, we use the retarded/advanced Nambu matrix functions 
$X(\omega)=[\omega^2/(1+\lambda^2)] M(\omega)/{\rm det} M(\omega)$, where
\begin{equation}
 M(\omega) =   \left( \begin{array}{cc} \omega^2-\alpha\Delta e^{-i\phi_0/2} & \alpha e^{i\phi_0/2} \sqrt{\Delta^2-\omega^2} \\ \alpha e^{-i\phi_0/2} \sqrt{\Delta^2-\omega^2} & \omega^2-\alpha\Delta e^{i\phi_0/2} \end{array}\right)
\end{equation}
with $\alpha=[2\lambda^2/(1+\lambda^2)]\Delta\cos(\phi_0/2)$. Using
the Andreev bound state energy $E_A$ in Eq.~\eqref{andreevstate}, we notice that ${\rm det} M = \omega^2 (\omega^2-E_A^2)$. As a consequence,  Eq.~\eqref{integraliphi0} leads to Eq.~\eqref{fracjos1} in the main text.

Next we discuss the function $Q(\omega_1,\omega_2)$ determining the finite-frequency noise $S_+(\omega)$, see Sec.~\ref{sec4b}.  To that end, we define the lead Nambu matrix $A(\omega)=[G^R-G^A](\omega)$ with $G^{R/A}$ in Eq.~\eqref{greensfunctions}, which corresponds to the spectral function.  We then obtain $Q(\omega_1,\omega_2)$  from Eq.~\eqref{noisegeneral} in the form
\begin{widetext}
\begin{equation}\label{QDE}
Q(\omega_1, \omega_2) = -\frac{1-\sqrt{1-\tau}}{1+\sqrt{1-\tau}} {\rm tr}_N\left [ A_{11}(\omega_1)
e^{i\sigma_z\phi_0/2} A_{22}(\omega_2) e^{-i\sigma_z\phi_0/2} - e^{i\sigma_z\phi_0/2} A_{21}(\omega_1) e^{i\sigma_z\phi_0/2} A_{21}(\omega_2)\right].
\end{equation}
Using Eq.~\eqref{greensfunctions}, the diagonal elements ($j=j'=1,2$) of the spectral function are given by the Nambu matrices
\begin{equation}
A_{jj}(\omega) = -i(1+\sqrt{1-\tau}) \left[ \frac{\pi}{2} [\delta(\omega-E_A)+\delta(\omega+E_A)] \left (\begin{array}{cc} \sqrt{\Delta^2-E_A^2} &e^{is_j\phi_0/2} \beta_j \\
 e^{-is_j\phi_0/2}\beta_j^\ast & \sqrt{\Delta^2-E_A^2} \end{array}\right) +\Theta(|\omega|-\Delta) \frac{|\omega|\sqrt{\omega^2- \Delta^2}}{\omega^2-E_A^2} \sigma_0\right],
\end{equation}
while for the off-diagonal component needed in Eq.~\eqref{QDE}, we obtain
 \begin{eqnarray} \nonumber
A_{21}(\omega) &=& -i(1+\sqrt{1-\tau}) e^{-i\sigma_z\phi_0/2}\Biggl[ \frac{\pi}{2} [\delta(\omega-E_A)-\delta(\omega+E_A)]\left (\begin{array}{cc} \beta_1^* &\sqrt{\Delta^2-E_A^2} e^{i\phi_0/2} \\
-\sqrt{\Delta^2-E_A^2}e^{-i\phi_0/2}& -\beta_1\end{array}\right) \\
&+& \Theta(|\omega|-\Delta) {\rm sgn}(\omega) \frac{E_A\sqrt{\omega^2-\Delta^2}}{\omega^2-E_A^2}\left (\begin{array}{cc} 0 & e^{i\phi_0/2} \\-e^{-i\phi_0/2}& 0\end{array}\right)\Biggr],
\end{eqnarray}
\end{widetext}
where  $s_{j}=(-1)^{j+1}$ and 
\begin{equation}
\beta_j=\left[ s_j \sqrt{1-\tau}\cos(\phi_0/2)-i\sin(\phi_0/2)\right]\Delta.
\end{equation}
Inserting these results into Eq.~\eqref{QDE}, we arrive at the expressions for $Q_{A-c}$ and $Q_{c-c}$ quoted in Eqs.~\eqref{QAc} and \eqref{Qcc}, respectively. 

\section{On S-TS junctions}  \label{appc}

In this appendix, we provide details about the calculation of the current through an S-TS junction, see Sec.~\ref{sec5}. The uncoupled GFs $\check{g}_{1,2}$ are then given by $\check{g}_1=\check{g}_S$ and $\check{g}_2=\check{g}_{TS}$.

First, we derive  the mean time-dependent current in Eq.~\eqref{currsts}. Let us start from Eq.~\eqref{currentgeneral}, which here takes the form
\begin{equation}\label{currentsts1}
I(t) = -{\rm Re}   \ {\rm tr}_N \left( W_{12}^\dagger(t) G^K_{12} (t,t) \right),
\end{equation}
with $W_{12}=\lambda e^{i\phi(t)/2}\Pi_\uparrow$. According to the Dyson equation \eqref{dyson}, we have 
\begin{equation}
\check G_{12}(t,t) = \int dt' \check g_{1}(t-t') W_{12}(t') \check{G}_{22}(t',t),  
\end{equation}
where $\check G^{-1}_{22}=\check g^{-1}_{2}-\check{\Sigma}$ involves the self-energy due to the tunnel coupling,
\begin{equation}\label{selfsts}
\check\Sigma(t,t')= \lambda^2 e^{-i[\phi(t)- \phi(t')]/2} \check\Sigma_{\rm eq} (t-t').
\end{equation}
Taking into account the reality constraint for the TS Nambu spinors, $\Psi=\sigma_x \Psi^\ast$, see Sec.~\ref{sec2a}, and using the projected  GF $\tilde g_1$ in Eq.~\eqref{projgf}, we find
\begin{eqnarray}\nonumber
\Sigma^{R/A}_{eq}(\omega) &=& \tilde g_1(\omega) - \sigma_x  \tilde g_1^{\ast} ( -\omega) \sigma_x ,\\ 
\Sigma_{eq}^K(\omega) &=& f(\omega) \left[ \Sigma_{eq}^R(\omega) - \Sigma_{eq}^A(\omega) \right].
\end{eqnarray}
Inserting the above expressions into Eq.~\eqref{currentsts1}, we arrive at Eq.~\eqref{currsts} in Sec.~\ref{sec5}.

Next, let us show that the equilibrium S-TS Josephson current for fixed phase difference $\phi_0$ vanishes identically in the absence of spin-flip tunneling.  For $V=0$, by employing the equilibrium relation for the Keldysh GF component \eqref{kcdef}, the 
Josephson current follows from Eq.~\eqref{currsts} as
\begin{eqnarray}\label{ists}
I(\phi_0) &=& -{\rm Re}\int \frac{d\omega}{2\pi} f(\omega) \left[ X_0^R(\omega)-X^A_0(\omega) \right],\\
X_0(\omega) &=& -\frac{\lambda^2 \omega}{\sqrt{\Delta_s^2-\omega^2}}  {\rm tr}_N \left(\Pi_\uparrow G_{22}(\omega) \right),
 \nonumber 
\end{eqnarray}
with
\begin{equation}
G_{22}(\omega) = \left( \left [g_{2}(\omega)\right]^{-1} + \frac{\lambda^2 \omega\sigma_0}{\sqrt{\Delta_s^2-\omega^2}}
\right)^{-1}.
\end{equation}
As a result, we obtain
\begin{equation}\label{X0RA}
X_0(\omega) = \frac{\lambda^2\omega^2 K(\omega)}{K^2(\omega)-\Delta^2(\Delta_s^2-\omega^2)},
\end{equation}
with $K(\omega)=\sqrt{(\Delta_s^2-\omega^2)(\Delta^2-\omega^2)}-\lambda^2\omega^2.$  Importantly, the $\phi_0$-dependence has dropped out completely, as can already be seen from Eq.~\eqref{currsts}.
By inserting Eq.~\eqref{X0RA} into Eq.~\eqref{ists}, we find $I(\phi_0)=0$, in accordance with  Ref.~\cite{Zazunov2012}.

Finally, starting from Eq.~\eqref{currsts}, we sketch the derivation of the expression (\ref{stsnv}) for the time-averaged (d.c.) current $I$ under a constant voltage bias.  Due to absence of MAR features,  one can effectively switch back to gauge I and work in the frequency representation.  Using $\omega_\pm=\omega\pm V$ and the projected self-energy components 
\begin{eqnarray}
\tilde \Sigma^{R/A}(\omega) &=& -\frac{\lambda^2\omega_-}{\sqrt{\Delta_s^2-\omega_-^2}}\Pi_\uparrow,\\
\nonumber
\tilde \Sigma^K(\omega) &=& -2i\lambda^2 f(\omega_-) \nu_1(\omega_-) \Pi_\uparrow,
\end{eqnarray}
we obtain the d.c.~current from Eq.~\eqref{currsts} as 
\begin{equation}
I= -\frac{e}{h}{\rm Re}\ {\rm tr}_N \int d\omega \left[ \tilde \Sigma^R(\omega)G_{22}^K(\omega)+\tilde \Sigma^K(\omega) G^A_{22}(\omega)\right].
\end{equation}
The retarded/advanced GF components  follow from the Dyson equation,
\begin{equation}
G_{22} (\omega) = \left( \left[g_{2}(\omega)\right]^{-1}-\Sigma(\omega)\right)^{-1},
\end{equation}
with the self-energy Nambu matrix
\begin{equation}  
\Sigma(\omega) = -\lambda^2 {\rm diag}\left( \frac{\omega_-}{\sqrt{\Delta_s^2-\omega_-^2}}, \frac{\omega_+}{\sqrt{\Delta_s^2-\omega_+^2}} \right),
\end{equation}
and 
\begin{eqnarray}
G_{22}^K(\omega) &=& G_{22}^R(\omega)\Bigl ( \Sigma^K(\omega)+\\ \nonumber 
&+& \left[g_{2}^R(\omega)\right]^{-1} g_{2}^K(\omega) \left[g_{2}^A(\omega)\right]^{-1} \Bigr) G^A_{22}(\omega),
\end{eqnarray}
where
\begin{equation}
\Sigma^K(\omega) = -2i\lambda^2 {\rm diag}\left[ f(\omega_-)\nu_1(\omega_-), f(\omega_+)\nu_1(\omega_+)\right].
\end{equation}
The above expressions yield Eq.~\eqref{stsnv} quoted in the main text.


\begin{thebibliography}{99}

\bibitem{Alicea2012}
J. Alicea, New directions in the pursuit of Majorana fermions in solid state systems, Rep. Prog. Phys. {\bf 75}, 076501 (2012).

\bibitem{Leijnse2012}
M. Leijnse and K. Flensberg, Introduction to topological superconductivity and Majorana fermions, Semicond. Sci. Techn. {\bf 27}, 124003 (2012).

\bibitem{Beenakker2013}
C.W.J. Beenakker, Search for Majorana fermions in superconductors, Annu. Rev. Con. Mat. Phys. {\bf 4}, 113 (2013).

\bibitem{Franz2015}
S.R. Elliott and M. Franz, Majorana fermions in nuclear, particle, and solid-state physics, Rev. Mod. Phys. {\bf 87}, 137 (2015).

\bibitem{Beenakker2015}
C.W.J. Beenakker, Random-matrix theory of Majorana fermions and topological superconductors, Rev. Mod. Phys. {\bf 87}, 1037 (2015). 

\bibitem{Mourik2012}
V. Mourik, K. Zuo, S.M. Frolov, S.R. Plissard, E.P.A. Bakkers, and L.P. Kouwenhoven, Signatures of Majorana fermions in hybrid superconductor-semiconductor nanowire devices, Science {\bf 336}, 1003 (2012).

\bibitem{Das2012}
A. Das, Y. Ronen, Y. Most, Y. Oreg, M. Heiblum, and H. Shtrikman, Zero-bias peaks and splitting in an Al-InAs nanowire topological superconductor as a signature of Majorana fermions, Nat. Phys.  {\bf 8}, 887 (2012).

\bibitem{Churchill2013}
H.O.H. Churchill, V. Fatemi, K. Grove-Rasmussen, M.T. Deng, P. Caroff, H.Q. Xu, and C.M. Marcus, Superconductor-nanowire devices from tunneling to the multichannel regime: Zero-bias oscillations and magnetoconductance crossover,
Phys. Rev. B {\bf 87}, 241401(R) (2013).

\bibitem{Albrecht2016} 
S.M. Albrecht, A.P. Higginbotham, M. Madsen, F. Kuemmeth, T.S. Jespersen, J. Nyg{\aa}rd, P. Krogstrup, and C.M. Marcus, Exponential Protection of Zero Modes in Majorana Islands, Nature {\bf 531}, 206 (2016).

\bibitem{Yazdani2014}
S. Nadj-Perge, I.K. Drozdov, J. Li, H. Chen, S. Jeon, J. Seo, A.H. MacDonald, B.A. Bernevig, and A. Yazdani, Observation of Majorana fermions in ferromagnetic atomic chains on a superconductor, Science {\bf 346}, 602 (2014).

\bibitem{Franke2015}
M. Ruby, F. Pientka, Y. Peng, F. von Oppen, B.W. Heinrich, and K.J. Franke,  End States and Subgap Structure in Proximity-Coupled Chains of Magnetic Adatoms, Phys. Rev. Lett. {\bf 115}, 197204 (2015).

\bibitem{BTK1982}
G.E. Blonder, M. Tinkham, and T.M. Klapwijk, Transition from metallic to tunneling regimes in superconducting microconstrictions: Excess current, charge imbalance, and supercurrent conversion, Pys. Rev. B {\bf 25}, 4515 (1982).

\bibitem{Nazarov2009}
Yu.V. Nazarov and Ya.M. Blanter, \textit{Quantum Transport: Introduction to Nanoscience} (Cambridge University Press, Cambridge, UK, 2010).

\bibitem{Cuevas1996}
 J.C. Cuevas, A. Mart{\'i}n-Rodero, and A. Levy Yeyati, Hamiltonian approach to the transport properties of superconducting quantum point contacts, Phys. Rev. B {\bf 54}, 7366 (1996).

\bibitem{ALY1997}
A. Levy Yeyati, J. C. Cuevas, A. L{\'o}pez-D{\'a}valos, and A. Mart{\'i}n-Rodero, Resonant tunneling through a small quantum dot coupled to superconducting leads, Phys. Rev. B {\bf 55}, R6137 (1997).

\bibitem{Zazunov2006}
A. Zazunov, R. Egger, C. Mora, and T. Martin, Superconducting transport through a vibrating molecule, Phys. Rev. B {\bf 73}, 214501 (2006).

\bibitem{Sun1999}
Q. Sun, J. Wang, and T. Lin, Photon-assisted Andreev tunneling through a mesoscopic hybrid system, Phys. Rev. B {\bf 59}, 13126 (1999). 

\bibitem{Bergeret2005}
F.S. Bergeret, A. Levy Yeyati, and A. Mart{\'i}n-Rodero, Inverse proximity effect in superconductor-ferromagnet structures: From the ballistic to the diffusive limit, Phys. Rev. B {\bf 72}, 064524 (2005).

\bibitem{ALY2003}
A. Levy Yeyati, A. Mart{\'i}n-Rodero, and E. Vecino, Nonequilibrium Dynamics of Andreev States in the Kondo Regime, Phys. Rev. Lett. {\bf 91}, 266802 (2003).

\bibitem{ALY2005}
A. Levy Yeyati, J.C. Cuevas, and A. Mart{\'i}n-Rodero, 
Dynamical Coulomb Blockade of Multiple Andreev Reflections,
Phys. Rev. Lett. {\bf 95}, 056804 (2005).

\bibitem{Melin2002}
R. M{\'e}lin and D. Feinberg, Transport theory of multiterminal hybrid structures, Eur. Phys. J B {\bf 26}, 101 (2002).

\bibitem{Datta1998}
M.P. Samanta and S. Datta, Electrical transport in junctions between unconventional superconductors: Application of the Green's-function formalism,
Phys. Rev. B {\bf 57}, 10972 (1998).

\bibitem{Sengupta2001}
K. Sengupta, Igor \u{Z}uti{\'c}, H.J. Kwon, V.M. Yakovenko, and S. Das Sarma, Midgap edge states and pairing symmetry of quasi-one-dimensional organic superconductors,
Phys. Rev. B {\bf 63}, 144531 (2001). 

\bibitem{Cuevas2001}
J. Cuevas and M. Fogelstr\"om,
Quasiclassical description of transport through superconducting contacts,
Phys. Rev. B {\bf 64}, 104502 (2001). 

\bibitem{Flensberg2010}
K. Flensberg, Tunneling characteristics of a chain of Majorana bound states, Phys. Rev. B {\bf 82}, 180516(R) (2010).

\bibitem{Leijnse2011} 
M. Leijnse and K. Flensberg, Scheme to measure Majorana fermion lifetimes using a quantum dot, Phys. Rev. B {\bf 84}, 140501(R) (2011).

\bibitem{Liu2011}
D.E. Liu and H.U. Baranger, Detecting a Majorana-fermion zero mode using a quantum dot, Phys. Rev. B {\bf 84}, 201308(R) (2011).
 
\bibitem{Kitaev2001}
A.Yu. Kitaev, Unpaired Majorana fermions in quantum wires, Usp. Fiz. Nauk (Suppl) {\bf 171}, 131 (2001).

\bibitem{Komnik2015}
A. Komnik, Transport properties of hybrid topological superconductor devices in contact with an environment, Phys. Rev. B {\bf 93}, 125117 (2016).

\bibitem{Law2009}
K.T. Law, P.A. Lee, and T.K. Ng, Majorana Fermion Induced Resonant Andreev Reflection, Phys. Rev. Lett. {\bf 103}, 237001 (2009).

\bibitem{Wimmer2011}
M. Wimmer, A.R. Akhmerov, J.P. Dahlhaus, and C.W.J. Beenakker, Quantum point contact as a probe of a topological superconductor, New J. Phys. {\bf 13}, 053016 (2011).

\bibitem{Prada2012}
E. Prada, P. San-Jose, and R. Aguado, Transport spectroscopy of $NS$ nanowire junctions with Majorana fermions, Phys. Rev. B {\bf 86}, 180503(R) (2012).

\bibitem{Bolech2007}
C.J. Bolech and E. Demler, Observing Majorana bound States in $p$-Wave Superconductors Using Noise Measurements in Tunneling Experiments,
Phys. Rev. Lett.  {\bf 98}, 237002 (2007)

\bibitem{Nilsson2008}
J. Nilsson,  A.R. Akhmerov, and C.W.J. Beenakker, Splitting of a Cooper Pair by a Pair of Majorana Bound States, Phys. Rev. Lett. {\bf 101}, 120403 (2008).

\bibitem{Golub2011}
A. Golub and B. Horowitz, Shot noise in a Majorana fermion chain, Phys. Rev. B {\bf 83}, 153415 (2011).

\bibitem{FuKane2009}
L. Fu and C.L. Kane, Josephson current and noise at a superconductor/quantum-spin-Hall-insulator/superconductor junction, Phys. Rev. B {\bf 79}, 161408(R) (2009).

\bibitem{Jiang2011}
L. Jiang, D. Pekker, J. Alicea, G. Refael, Y. Oreg, and F. von Oppen, Unconventional Josephson signatures of Majorana Bound States, Phys. Rev. Lett. {\bf 107}, 236401 (2011).

\bibitem{Virtanen2013}
P. Virtanen and P. Recher, Microwave spectroscopy of Josephson junctions in topological superconductors, Phys. Rev. B {\bf 88}, 144507 (2013).

\bibitem{Belzig2015}
J.I. V{\"a}yrynen, G. Rastelli, W. Belzig, and L.I. Glazman, Microwave signatures of Majorana states in a topological Josephson junction, Phys. Rev. B {\bf 92}, 134508 (2015).

\bibitem{Badiane2011}
D. Badiane, M. Houzet, and J.S. Meyer, Nonequilibrium Josephson Effect through Helical Edge States, Phys. Rev. Lett. {\bf 107}, 177002 (2011).

\bibitem{Houzet2013}
M. Houzet, J.S. Meyer, D.M. Badiane, and L.I. Glazman, Dynamics of Majorana States in a Topological Josephson Junction, Phys. Rev. Lett. {\bf 111}, 046401 (2013).

\bibitem{Aguado2013}
P. San-Jose, J. Cayao, E. Prada, and R. Aguado, Multiple Andreev reflection and critical current in topological superconducting nanowire junctions, New J. Phys. {\bf 15}, 075019 (2013).

\bibitem{Zazunov2012} 
A. Zazunov and R. Egger, Supercurrent blockade in Josephson junctions with a Majorana wire, Phys. Rev.  B {\bf 85}, 104514 (2012).

\bibitem{Ioselevich2015}
P.A. Ioselevich, P.M. Ostrovsky, and M.V. Feigelman, Josephson current between topological and conventional superconductors,  Phys. Rev. B {\bf 93}, 125435 (2016).

\bibitem{Peng2015}
Y. Peng, F. Pientka, Y. Vinkler-Aviv, L.I. Glazman, and F. von Oppen, Robust Majorana Conductance Peaks for a Superconducting Lead, Phys. Rev. Lett. {\bf 115}, 266804 (2015).

\bibitem{Loss2013}
D. Rainis, L. Trifunovic, J. Klinovaja, and D. Loss, Towards a realistic transport modeling in a superconducting nanowire with Majorana fermions, Phys. Rev. B {\bf 87}, 024515 (2013).

\bibitem{Avishai2001}
Y. Avishai, A. Golub, and A.D. Zaikin, Tunneling through an Anderson impurity between superconductors, Phys. Rev. B {\bf 63}, 134515 (2001).


\bibitem{footnote2}
To see this, we employ a canonical transformation to new operators, $\tilde c_\uparrow=\cos(\theta)c_{1,\uparrow}+\sin(\theta)c_{1,\downarrow}$ and
$\tilde c_\downarrow=-\sin(\theta)c_{1,\uparrow}+\cos(\theta)c_{1,\downarrow}$. This transformation leaves the GF (\ref{gfs}) invariant, but now only $\tilde c_\uparrow$ is tunnel-coupled to the TS wire, while its time-reversed partner $\tilde c_{\downarrow}$ remains decoupled from it. (The pairing term $\sim\Delta_s$ will still  couple $\tilde c_\uparrow$ and $\tilde c_\downarrow$ within the bulk of the BCS superconductor.)  Without loss of generality, we can therefore put $\theta=0$. 


\bibitem{Tanaka2000}
S. Kashiwaya and Y. Tanaka,
Tunnelling effects on surface bound states in unconventional superconductors,
Rep. Prog. Phys. {\bf 63}, 1641 (2000).

\bibitem{Khlus1987}
V.A. Khlus, Current and voltage fluctuations in microjunctions between normal metals
and superconductors, Zh. Eksp. Teor. Fiz. {\bf 93}, 2179 (1987) [Sov. Phys. JETP {\bf 66}, 1243 (1987)].

\bibitem{Zaitsev1980}
A.V. Zaitsev, Theory of pure short S-c-S and S-c-N microjunctions,
Zh. Eksp. Teor. Fiz. {\bf 78}, 221 (1980) [Sov. Phys. JETP {\bf 51}, 111 (1980)].


\bibitem{ALY1996}
A. Mart{\'i}n-Rodero, A. Levy Yeyati, and F.J. Garc{\'i}a-Vidal, Thermal noise in superconducting quantum point contacts,
Phys. Rev. B {\bf 53}, R8891(R) (1996).

\bibitem{AZ2005}
A. Zazunov, V. S. Shumeiko, G. Wendin, and E. N. Bratus',
Dynamics and phonon-induced decoherence of Andreev level qubit,
Phys. Rev. B {\bf 71}, 214505 (2005).

\bibitem{Kos2013}
F. Kos, S.E. Nigg, and L.I. Glazman, Frequency-dependent admittance of a short superconducting weak link, Phys. Rev. B {\bf 87}, 174521 (2013).

\bibitem{Andersen2011}
B.M. Andersen, K. Flensberg, V. Koerting, and J. Paaske, Nonequilibrium Transport through a Spinful Quantum Dot with Superconducting Leads, Phys. Rev. Lett. {\bf 107}, 256802 (2011). 

\bibitem{Alicea2011}
J. Alicea, Y. Oreg, G. Refael, F. von Oppen, and M.P.A. Fisher, Non-Abelian statistics and topological quantum information processing in 1D wire networks, Nat. Phys. {\bf 7}, 412 (2011).

\bibitem{Weithofer2014}
L. Weithofer, P. Recher, and T.L. Schmidt, Electron transport in multiterminal networks of Majorana bound states, Phys. Rev. B {\bf 90}, 205416 (2014).

\bibitem{Tarasinski2015}
B. Tarasinski, D. Chevallier, J.A. Hutasoit,  B. Baxevanis, and C.W.J. Beenakker, Quench dynamics of fermion-parity switches in a Josephson junction, Phys. Rev. B {\bf 92}, 144306 (2015).

\bibitem{Haim2015}
A. Haim, E. Berg, F. von Oppen, and Y. Oreg, Signatures of Majorana Zero Modes in Spin-Resolved Current Correlations, Phys. Rev. Lett. {\bf 114}, 166406 (2015).

\bibitem{DasSarma2012}
S. Das Sarma, J.D. Sau, and T.D. Stanescu, Splitting of the zero-bias conductance peak as smoking  gun evidence for the existence of a Majorana mode in a superconductor-semiconductor nanowire, Phys. Rev. B {\bf 86}, 220506(R) (2012). 

\bibitem{Aguado2012}
P. San-Jose, E. Prada, and R. Aguado, ac Josephson Effect in Finite-Length Nanowire Junctions with Majorana Modes, Phys. Rev. Lett. {\bf 108}, 257001 (2012).

\bibitem{Fu2010}
L. Fu, Electron teleportation via Majorana bound states in a mesoscopic superconductor, Phys. Rev. Lett. {\bf 104}, 056402 (2010).

\bibitem{Zazunov2011}
A. Zazunov, A. Levy Yeyati, and R. Egger, Coulomb blockade of Majorana-fermion-induced transport, Phys. Rev. B {\bf 84}, 165440 (2011).

\bibitem{BeriCooper2012}
B. B{\'e}ri and N. Cooper, Topological Kondo effect with Majorana fermions, Phys. Rev. Lett. {\bf 109}, 156803 (2012).

\bibitem{Huetzen2012} 
R. H\"utzen, A. Zazunov, B. Braunecker, A.L. Yeyati, and R. Egger, Majorana Single-Charge Transistor, Phys. Rev. Lett. {\bf 109}, 166403 (2012).

\bibitem{arrachea}
L. Arrachea, G.S. Lozano, and A.A. Aligia, Thermal transport in one-dimensional spin heterostructures, Phys. Rev. B {\bf 80}, 014425 (2009).

\end{thebibliography}
\end{document}